\documentclass[aps,twocolumn,preprintnumbers,amsmath,amssymb,nofootinbib,superscriptaddress,notitlepage]{revtex4}
\usepackage{graphicx,color,dcolumn,booktabs,bm}
\usepackage{longtable,lscape}
\usepackage{txfonts}
\usepackage{overpic}
\usepackage{amssymb}
\usepackage{indentfirst}
\usepackage{feynmf}
\usepackage{slashed}
\usepackage{cases}
\usepackage{multirow}
\usepackage{pifont}
\usepackage{appendix}
\usepackage[figuresright]{rotating}
\usepackage[colorlinks,
            citecolor=blue,
            anchorcolor=red,
            menucolor=red,
            linkcolor=red,
            filecolor=red,
            runcolor=red,
            urlcolor=blue,
            frenchlinks=red]{hyperref}
\usepackage{epstopdf}
\usepackage{bm}
\usepackage{tabularx}
\usepackage{threeparttable}
\usepackage{bbding}
\usepackage{array}
\usepackage[section]{placeins}
\usepackage{makecell}
\usepackage{comment}
\makeatletter

\newcommand{\Rmnum}[1]{\expandafter\@slowromancap\romannumeral #1@}
\makeatother

\begin{document}
\title{Role of electromagnetic interaction in the $X(3872)$ and its analogs}

\author{Ping Chen}
\affiliation{School of Physical Science and Technology, Lanzhou University, Lanzhou 730000, China}
\affiliation{Research Center for Hadron and CSR Physics, Lanzhou University and Institute of Modern Physics of CAS, Lanzhou 730000, China}

\author{Zhan-Wei Liu}\email{liuzhanwei@lzu.edu.cn}
\affiliation{School of Physical Science and Technology, Lanzhou University, Lanzhou 730000, China}
\affiliation{Research Center for Hadron and CSR Physics, Lanzhou University and Institute of Modern Physics of CAS, Lanzhou 730000, China}
\affiliation{Lanzhou Center for Theoretical Physics, Key Laboratory of Theoretical Physics of Gansu Province, Key Laboratory of Quantum Theory and Applications of MoE, and MoE Frontiers Science Center for Rare Isotopes, Lanzhou University, Lanzhou 730000, China}

\author{Zi-Le Zhang}
\affiliation{School of Physical Science and Technology, Lanzhou University, Lanzhou 730000, China}
\affiliation{Research Center for Hadron and CSR Physics, Lanzhou University and Institute of Modern Physics of CAS, Lanzhou 730000, China}

\author{Si-Qiang Luo}\email{luosq15@lzu.edu.cn}
\affiliation{School of Physical Science and Technology, Lanzhou University, Lanzhou 730000, China}
\affiliation{Research Center for Hadron and CSR Physics, Lanzhou University and Institute of Modern Physics of CAS, Lanzhou 730000, China}
\affiliation{Lanzhou Center for Theoretical Physics, Key Laboratory of Theoretical Physics of Gansu Province, Key Laboratory of Quantum Theory and Applications of MoE, and MoE Frontiers Science Center for Rare Isotopes, Lanzhou University, Lanzhou 730000, China}

\author{Fu-Lai Wang}\email{wangfl2016@lzu.edu.cn}
\affiliation{School of Physical Science and Technology, Lanzhou University, Lanzhou 730000, China}
\affiliation{Research Center for Hadron and CSR Physics, Lanzhou University and Institute of Modern Physics of CAS, Lanzhou 730000, China}
\affiliation{Lanzhou Center for Theoretical Physics, Key Laboratory of Theoretical Physics of Gansu Province, Key Laboratory of Quantum Theory and Applications of MoE, and MoE Frontiers Science Center for Rare Isotopes, Lanzhou University, Lanzhou 730000, China}
\author{Jun-Zhang Wang}\email{wangjzh2022@pku.edu.cn}
\affiliation{School of Physics and Center of High Energy Physics, Peking University, Beijing 100871, China\\}

\author{Xiang Liu}\email{xiangliu@lzu.edu.cn}
\affiliation{School of Physical Science and Technology, Lanzhou University, Lanzhou 730000, China}
\affiliation{Research Center for Hadron and CSR Physics, Lanzhou University and Institute of Modern Physics of CAS, Lanzhou 730000, China}
\affiliation{Lanzhou Center for Theoretical Physics, Key Laboratory of Theoretical Physics of Gansu Province, Key Laboratory of Quantum Theory and Applications of MoE, and MoE Frontiers Science Center for Rare Isotopes, Lanzhou University, Lanzhou 730000, China}

\begin{abstract}
We investigate the role of the electromagnetic interaction in the formation and decay of the $X(3872)$. The binding properties of the $X(3872)$ are studied by assuming the molecular nature and considering the $S$-$D$ wave mixing, isospin breaking, and coupled channel effects, and in particular the correction from the electromagnetic interaction. The radiative decays can better reflect the difference between the charged and neutral $D\bar D^*$ components, since the electromagnetic interaction explicitly breaks the isospin symmetry. We further study the radiative decay widths with the obtained wave functions for different $D\bar D^*$ channels. We also explore other similar hidden-charm molecular states. The electromagnetic interaction can make the molecule tighter. Our result of the radiative decay width for $X(3872)\rightarrow \gamma J/\psi$ is in agreement with the experiment. The branching ratio $R_{\gamma\psi}$ is less than 1 in our framework, which supports the Belle and BES\Rmnum{3} measurements.
\end{abstract}

\affiliation{}

\maketitle
\section{Introduction}\label{introduction}
The study of hadron spectroscopy can provide a guidance to our understanding of nonperturbative behavior of the strong interaction. Until now, great progress has been made through the construction of the conventional hadron family. Since 2003, we have entered new phase in the exploration of hadron spectroscopy. A starting point is the observation of the $X(3872)$ by the Belle Collaboration in $B^\pm \rightarrow K^\pm\pi^+\pi^- J/\psi$ \cite{Belle:2003nnu}. There is a low-mass puzzle here, since it appears too light to be the charmonium $\chi_{c1}(2P)$ \cite{Godfrey:1985xj,Capstick:1986ter}. Subsequently, many exotic hadron states are discovered by several experimental collaborations around the world, such as the hidden-charm pentaquarks $P_c(4380)$ \cite{LHCb:2015yax}, $P_c(4312)$, $P_c(4440)$, $P_c(4457)$ \cite{LHCb:2019kea}, $P_{cs}(4459)$ \cite{LHCb:2020jpq}, $P_{\psi s}^{\Lambda}(4338)$ \cite{LHCb:2022ogu}, the double-charm tetraquark $T_{cc}^+(3875)$ \cite{LHCb:2021vvq} and abundant charmoniumlike $XYZ$ states (see reviews \cite{Swanson:2006st,Chen:2016qju,Chen:2016spr,Esposito:2016noz,Olsen:2017bmm,Brambilla:2019esw,Liu:2019zoy,Chen:2022asf}). The observation of these new hadronic states provides important insights into the construction of the 2.0 version of the hadron family and extends our knowledge of the matter world.

As the first charmoniumlike state, the $X(3872)$ has been studied for over 20 years, but we still do not fully understand its nature. So far, many theoretical approaches have been proposed to unravel the nature of the $X(3872)$, which is explained as the $D\bar{D}^*$ hadronic molecular state \cite{Tornqvist:1993vu,Swanson:2003tb,Wong:2003xk,Close:2003sg,Voloshin:2003nt,Tornqvist:2004qy,AlFiky:2005jd,Thomas:2008ja,Liu:2008fh,Liu:2009qhy,Lee:2009hy,Braaten:2010mg,Wang:2013kva,Baru:2013rta,Baru:2015nea,Song:2023pdq,Gamermann:2009uq}, tetraquark \cite{Maiani:2004vq,Ebert:2005nc,Hogaasen:2005jv,Barnea:2006sd,Matheus:2006xi,Vijande:2007fc,Wang:2013vex}, charmonium \cite{Barnes:2003vb,Eichten:2004uh,Suzuki:2005ha,Kong:2006ni,Voloshin:2007dx,Kalashnikova:2010hv,Ferretti:2013faa,Ferretti:2014xqa}, hybrid \cite{Li:2004sta}, and  a mixture of a charmonium with a $D\bar{D}^*$ component \cite{Meng:2005er,Meng:2014ota}. The exotic nature of the $X(3872)$ is embodied in the mass and narrow width with $m_X=3871.65\pm0.06$ MeV and $\Gamma_X=1.19\pm0.21$ MeV as collected in the Particle Data Group (PDG) \cite{ParticleDataGroup:2022pth}. The hadronic molecular state becomes a plausible explanation for the extreme proximity of the $X(3872)$ to the $D^0\bar{D}^{*0}$ threshold. Its narrow width is not easily accommodated in the conventional charmonium picture. In addition, Baru \textit{et al}. \cite{Baru:2016iwj} predicted the existence of three degenerate spin partners of the $X(3872)$ with quantum numbers $0^{++}$, $1^{+-}$ and $2^{++}$. For the negative $C$-parity partner of $X(3872)$, there is some experimental evidence for such a negative $C$-parity state, named as $\tilde{X}(3872)$, reported by the COMPASS Collaboration \cite{COMPASS:2017wql}. Of course, it should be clarified in the future with the accumulation of higher precision data. For the isoscalar $D\bar{D}$ with $J^{PC}=0^{++}$ bound state, no clear evidence has yet been found though there are some attempts \cite{Gamermann:2007mu,Dai:2020yfu,Wang:2020elp} to extract such a state from the available experimental date \cite{Belle:2005rte,Belle:2007woe,BaBar:2010jfn}. This could be because no easily detectable decay modes are available since its mass is below the $D\bar{D}$ threshold. As for the $2^{++}$ $D^*\bar{D}^*$ bound state, no evidence has been found yet. One possible reason is that the coupling to ordinary charmonia could either move the $2^{++}$ pole deep into the complex energy plane and thus make it invisible \cite{Cincioglu:2016fkm} or make the $D^*\bar{D}^*$ interaction  unbound in the $2^{++}$ sector \cite{Ortega:2020tng,Ortega:2017qmg}.  

In the process of studying the hadronic molecular states, many contributions including the $S$-$D$ wave mixing effect \cite{Swanson:2003tb,Li:2012cs,Wang:2013kva}, the coupled channel effect \cite{Swanson:2003tb,Liu:2009qhy,Li:2012cs}, the isospin breaking effect  \cite{Close:2003sg,Tornqvist:2004qy,Thomas:2008ja,Lee:2009hy,Li:2012cs,Gamermann:2009fv,Takizawa:2010rxa} and the recoil correction effect \cite{Zhao:2014gqa} are introduced to decode the nature of the $X(3872)$. This can provide the important information for predicting the binding properties of bound states. 

In addition to the above effects related to the strong interaction, the electromagnetic interaction can also play a role in the formation of the $X(3872)$, since it is very close to the $D\bar D^*$ threshold. The role of the electromagnetic interaction in the formation of hadronic matter has been discussed in recent years. According to the lattice QCD study of the heavy dibaryons, the existence of a bound state in $\Omega_{ccc}\Omega_{ccc}$ can be broken by the Coulomb repulsion \cite{Lyu:2021qsh}. The $\Omega_{ccc}\Omega_{ccc}$ dibaryon with $J^P=0^+$ has the binding energy $E=-5.1$ MeV with the strong interaction alone, but is unbound when the Coulomb interaction is taken into account \cite{Liu:2021pdu}. The $\Omega_{bbb}\Omega_{bbb}$ dibaryon can be deeply bound by the strong interaction with the binding energy up to $-89(^{+16}_{-12})$ MeV, and the Coulomb repulsion can shift the strong binding by a few percent \cite{Mathur:2022nez}. The Coulomb interaction is not strong enough to break up the $DDD^*$ state \cite{Wu:2021kbu}.
Around the discussed $X(3872)$, the authors of Ref. \cite{Zhang:2020mpi} brought a new idea of the $X$ atom composed of $D^{\pm}D^{*\mp}$, which is formed mainly due to the Coulomb force. 

In this work, we still focus on the $X(3872)$ as a molecular-type $D\bar{D}^*$ state, mainly formed by the strong interaction, but the electromagnetic correction is taken into account. Under this consideration, the charged $D^{\pm}D^{*\mp}$ component
would be different from the neutral $D^{0}\bar D^{*0}/D^{*0}\bar D^{0}$ part. This difference can be accounted for by 
the radiative decays of the $X(3872)$. Obviously, this study can be related to the question of the inner structure of the $X(3872)$, which is currently an open question. 

There have been some theoretical studies of the radiative decay of $X(3872)$ under the assumptions that the $X(3872)$ is a hadronic molecular state \cite{Swanson:2004pp,Guo:2014taa}, charmonium \cite{Barnes:2003vb,Barnes:2005pb,Badalian:2012jz,Li:2009zu,Lahde:2002wj,Mehen:2011ds,Wang:2010ej,Yu:2023nxk}, and a $c\bar{c}-D\bar{D}^*$ mixing scheme \cite{Eichten:2005ga,Dong:2009uf,Cardoso:2014xda}.
As a typical work, Ref. \cite{Swanson:2004pp} obtained a small ratio $R_{\gamma\psi}$ in the molecular picture by light quark annihilation and the vector meson transition into a photon, where the authors suggested looking at the $X(3872)$ radiative decays into $\gamma J/\psi$ and $\gamma\psi(2S)$ as one of the promising tests for its molecular nature, which stimulates further experimental exploration of this issue. 

Until now, four experimental collaborations have announced their result of radiative decay of the $X(3872)$ into a $J/\psi$ or $\psi(2S)$. Here, the experimental ratio $R_{\gamma\psi}= \frac{\mathcal{B}[X(3872) \rightarrow \gamma\psi(2S)]}{\mathcal{B}[X(3872) \rightarrow \gamma J/\psi]}$ can be summarized as follows:
\begin{equation}
 R_{\gamma\psi}\left\{
     \begin{array}{lr}
        =3.4\pm1.4~(3.5\sigma)           &\rm{BaBar}\\
        =2.46\pm0.64\pm0.29~(4.4\sigma)   &\rm{LHCb}\\
        \textless~2.1~(90\%~\rm{C.L.})                    &\rm{Belle}\\
       \textless~0.59~(90\%~\rm{C.L.})                    & \rm{BES\Rmnum{3}} \nonumber
     \end{array}
\right..
\end{equation} 
It is obvious that the data from four different experiments appear to be inconsistent, which should be clarified in the near future with the accumulation of higher precision data. 
In fact, these measurement call into question the molecular assignment to the $X(3872)$ \cite{Swanson:2004pp}. Later, in Ref. \cite{Guo:2014taa}, the authors tried to provide a solution, i.e., to calculate the ratio $R_{\gamma\psi}$, the triangle loops and other diagrams were introduced. They claimed that the experimental ratio does not contradict
the calculated ratio if the $X(3872)$ is dominated by the $D\bar D^*$ hadronic molecule. 

As we all know, hadron spectroscopy represents the precision frontier of particle physics. 
With the promotion of precision of theoretical calculation, we are able to obtain more precise information about the mass spectrum and the corresponding spatial wave function of the $D\bar{D}^*$ molecular state associated with the $X(3872)$ and its partners. It makes us reconsider the radiative decay of the $X(3872)$
under this higher theoretical calculation precision, which will be main task of this work. 

In this work, we study the $X(3872)$ radiative decays by combining the wave function of the $D\bar D^*$ bound state and the scattering amplitudes of $D\bar D^*\to \gamma J/\psi$, $\gamma \psi(2S)$.
In addition, if the $X(3872)$ is a $D\bar{D}^*$ loosely bound state with the $J^{PC}=1^{++}$, there may exist the \textit{C}-parity counterpart with $J^{PC}=1^{+-}$. The $D_s\bar{D}^*_s$ could be bound by the SU(3) flavor symmetry \cite{Meng:2020cbk}. Reference \cite{Shi:2023mer} considered the radiative decay width for the $X(3872)$ partner $X_2$ with $J^{PC}=2^{++}$, treating the $X_2$ as a $D^*\bar{D}^*$ shallow bound state. 

Inspired by the above situation, in this work we investigate the role of the electromagnetic interaction in the $X(3872)$, the $D\bar{D}^*$ molecular state with $J^{PC}=1^{+-}$, and the $D_s\bar{D}^*_s$ molecular state with $J^{PC}=1^{+-}/1^{++}$. First, we calculate the corresponding bound state solutions including the electromagnetic interaction as well as the $S$-$D$ wave mixing effect, the coupled channel effect, and the isospin breaking effect. Then we discuss their radiative decays based on the obtained wave functions, which can provide the valuable information about the inner structure of hadrons. 
 
The rest of this paper is organized as follows. For the formation with the electromagnetic interaction, we present the framework in Sec. \ref{sec2} and the numerical results in Sec. \ref{sec3}. For the radiative processes, the effective Lagrangian approach is described in Sec. \ref{sec4} and our numerical results and discussions are shown in Sec. \ref{sec5}. A short summary follows in Sec. \ref{sec6}.

\section{Framework for formation with Coulomb interaction}\label{sec2}
This work considers the electromagnetic interaction effect in both the formation and the decay processes of the molecular states. For the binding properties in the formation, the Coulomb interaction itself would be sufficient at this stage. In this section, we include the Coulomb correction to study the $X(3872)$ as a possible $D\bar{D}^*$ molecular state as well as the $D\bar{D}^*$ ($1^{+-}$), $D_s\bar{D}_s^*$ ($1^{++}$) and $D_s\bar{D}_s^*$ ($1^{+-}$) molecular states. 
\setlength{\tabcolsep}{2.6mm}
\begin{table}[hbtp]\label{{notations}}
\centering
\caption{The shorthand notations of the eigenstates with the $C$-parity being $+1$ and $-1$ for $D\bar{D}^*$ and $D_s\bar D_s^*$ systems.}\label{notations}
\renewcommand\arraystretch{1.40}
\begin{tabular*}{85mm}{llc}
\toprule[1.0pt]
\toprule[1.0pt]
$J^{PC}$&{Notations}&\multicolumn{1}{c}{Configurations} \\
\midrule[0.75pt]
$1^{+ \pm}$
 &$[D^{0}\bar{D}^{*0}]$ & ${1\over\sqrt{2}}\left(D^{0}\bar{D}^{*0}\mp D^{*0}\bar{D}^{0}\right)$\\
~&$[D^+D^{*-}]$         & ${1\over\sqrt{2}}\left(D^+D^{*-}\mp D^{*+}D^-\right)$\\
~&$[D\bar{D}^*]$        & ${1\over2}\left[\left(D^{0}\bar{D}^{*0}\mp D^{*0}\bar{D}^{0}\right)+\left(D^{+}D^{*-}\mp D^{*+}D^{-}\right)\right]$\\
~&$[D_s\bar{D}_s^*]$ & $\frac{1}{\sqrt{2}}(D_s^+D_s^{*-}\mp D_s^-D_s^{*+})$\\
\bottomrule[1.0pt]
\bottomrule[1.0pt]
\end{tabular*}
\end{table}
\subsection{The flavor and spin-orbital wave functions}\label{flavor wave function}
\setlength{\tabcolsep}{0.7mm}
\begin{table*}[hbtp]
\renewcommand{\arraystretch}{1.0} \caption{The different channels for Cases \Rmnum{1}, \Rmnum{2}, and \Rmnum{3} of the $X(3872)$ with $J^{PC}=1^{++}$ and its \textit{C}-parity partner with $J^{PC}=1^{+-}$. Shorthand notations are used in Table \ref{notations}, and the superscript $\mathcal{C}$ indicates that the Coulomb interaction between the charged mesons is added. We consider the isospin symmetry without Coulomb interaction in Case \Rmnum{1}, include the isospin breaking effect due to meson mass difference but still no Coulomb interaction in Case \Rmnum{2}, and consider both the isospin breaking effect and Coulomb interaction into account in Case \Rmnum{3}.}\label{Channel}
\renewcommand\arraystretch{1.15}
\begin{tabular*}{18cm}{@{\extracolsep{\fill}}cclllllllll}
\toprule[1.00pt]
\toprule[1.00pt]

      \multirow{2}{*}{$J^{PC}$}&\multirow{2}{*}{Cases}       & \multicolumn{8}{c}{Channels}\\
      \Xcline{3-10}{0.75pt}
      & & \multicolumn{1}{c}{1} & \multicolumn{1}{c}{2} & \multicolumn{1}{c}{3} & \multicolumn{1}{c}{4} & \multicolumn{1}{c}{5} & \multicolumn{1}{c}{6} & \multicolumn{1}{c}{7} & \multicolumn{1}{c}{8}\\
\midrule[0.75pt]
$1^{++}$&              \Rmnum{1}     &$[D\bar{D}^{*}]|^3S_1\rangle$&
              $[D\bar{D}^{*}]|^3D_1\rangle$&
              \multicolumn{1}{c}{}&
              \multicolumn{1}{c}{}&
              \multicolumn{1}{c}{}&
              \multicolumn{1}{c}{}&
              \multicolumn{1}{c}{}&
              \multicolumn{1}{c}{}\\ [3pt]
~&\Rmnum{2}  &$[D^0\bar{D}^{*0}]|^3S_1\rangle$&
              $[D^0\bar{D}^{*0}]|^3D_1\rangle$&
              $[D^+D^{*-}]|^3S_1\rangle$&
              $[D^+D^{*-}]|^3D_1\rangle$&
              \multicolumn{1}{c}{}&
              \multicolumn{1}{c}{}&
              \multicolumn{1}{c}{}&
              \multicolumn{1}{c}{} \\ [3pt]
~&\Rmnum{3}  &$[D^0\bar{D}^{*0}]|^3S_1\rangle$&
              $[D^0\bar{D}^{*0}]|^3D_1\rangle$&
              $[D^+D^{*-}]|^3S_1\rangle^{\mathcal{C}}$&
              $[D^+D^{*-}]|^3D_1\rangle^{\mathcal{C}}$&
              \multicolumn{1}{c}{}&
              \multicolumn{1}{c}{}&
              \multicolumn{1}{c}{}&
              \multicolumn{1}{c}{}\\[3pt]
\addlinespace[0.45em]
$1^{+-}$&              \Rmnum{1}    &$[D\bar{D}^{*}]|^3S_1\rangle$&
             $[D\bar{D}^{*}]|^3D_1\rangle$&
             $[D^*\bar{D}^*]|^3S_1\rangle$&
             $[D^*\bar{D}^*]|^3D_1\rangle$&
             \multicolumn{1}{c}{}&
             \multicolumn{1}{c}{}&
             \multicolumn{1}{c}{}&
             \multicolumn{1}{c}{}\\[3pt]
~&\Rmnum{2} &$[D^0\bar{D}^{*0}]|^3S_1\rangle$&
             $[D^0\bar{D}^{*0}]|^3D_1\rangle$&
             $[D^+D^{*-}]|^3S_1\rangle$&
             $[D^+D^{*-}]|^3D_1\rangle$&
             $[D^{*0}\bar{D}^{*0}]|^3S_1\rangle$&
             $[D^{*0}\bar{D}^{*0}]|^3D_1\rangle$&
             $[D^{*+}\bar{D}^{*-}]|^3S_1\rangle$&
             $[D^{*+}\bar{D}^{*-}]|^3D_1\rangle$\\ [3pt]

~&\Rmnum{3} &$[D^0\bar{D}^{*0}]|^3S_1\rangle$&
             $[D^0\bar{D}^{*0}]|^3D_1\rangle$&
             $[D^+D^{*-}]|^3S_1\rangle^{\mathcal{C}}$&
             $[D^+D^{*-}]|^3D_1\rangle^{\mathcal{C}}$&
             $[D^{*0}\bar{D}^{*0}]|^3S_1\rangle$&
             $[D^{*0}\bar{D}^{*0}]|^3D_1\rangle$&
             $[D^{*+}\bar{D}^{*-}]|^3S_1\rangle^{\mathcal{C}}$&
             $[D^{*+}\bar{D}^{*-}]|^3D_1\rangle^{\mathcal{C}}$\\ [3pt]
\toprule[1.00pt]
\toprule[1.00pt]
\end{tabular*}
\end{table*}
The mass of the $X(3872)$ is very close to the $D^0\bar{D}^{*0}$ threshold, which shows the possibility of the $X(3872)$ as a molecular state. Since the mass difference between the $D^0\bar{D}^{*0}$ and $D^+D^{*-}$ channels is about 8 MeV, there is a strong coupling between them. In addition, the quantum number of the $X(3872)$ is $J^{PC}=1^{++}$ \cite{Belle:2011wdj,LHCb:2013kgk}, so we need to construct the $C$-parity eigenstates. Here, the charmed mesons $D$ and ${D}^*$ satisfy $C D C^{-1}=\bar{D}$ and $\textit{C}D^*\textit{C}^{-1}=-\bar{D}^*$ under the charge conjugate transformation \cite{Grinstein:1992qt,Cincioglu:2016fkm}, respectively. We summarize the \textit{C}-parity $D\bar{D}^*$ eigenstates and use the shorthand notations in Table \ref{notations}.

For the $D_{(s)}^{(*)}\bar{D}_{(s)}^*$ systems discussed, the corresponding spin-orbit wave functions are as follows
\begin{eqnarray}
D_{(s)}\bar{D}_{(s)}^*&:&|\,{}^{2S+1}L_{J}\rangle\sim\sum_{m,m_L}
C^{J,M}_{1m,Lm_L}\epsilon^{\mu}_{m}\,Y_{L,m_L},\\
{D}^*\bar{D}^{*}&:&|\,{}^{2S+1}L_{J}\rangle\sim\!\!\!\!\!\sum_{m_1,m_2,m_L}\!\!\!\!\!
C^{S,m_S}_{1m_1,1m_2}C^{J,M}_{Sm_S,Lm_L}\epsilon^{\mu}_{m_1}\epsilon^{\nu}_{m_2}
\,Y_{L,m_L},
\end{eqnarray}
where $C^{j,m}_{j_1m_1,j_2m_2}$ is the Clebsch-Gordan coefficient, and $Y_{L,m_L}$ is spherical harmonics function.

The electromagnetic interaction plays a similar role to the quark mass difference in the neutron-proton mass difference. We therefore consider the $X(3872)$ as the possible $D\bar{D}^*$ molecular state in three different scenarios. The exact $D\bar{D}^*$ isospin singlet is assumed in Case \Rmnum{1}. The mass splitting between the charged and neutral $D^{(*)}$ mesons is taken into account for Case \Rmnum{2}, and we further introduce the Coulomb interaction between the charged pair in Case \Rmnum{3}. 
Similarly, we can also construct three scenarios for the \textit{C}-parity counterpart. We summarize three different cases for the possible $D\bar{D}^*$ molecular states with $J^{PC}=1^{++}$ and $J^{PC}=1^{+-}$ in Table \ref{Channel}. The $D^*\bar{D}^*$ channel is negligible for the $J^{PC}=1^{++}$ sector as it is only allowed in a relative $|^5D_1\rangle$ partial wave, but this channel has a significant effect through the $S$-wave interaction for the $J^{PC}=1^{+-}$ state. 

Since there is no neutral $D_s$ meson, we study the $D_s\bar{D}_s^*$ molecular states only for two cases, with and without the Coulomb interaction.

\subsection{The interaction potentials and Schr\"{o}dinger equation}\label{Lagrangian}

The interaction potentials between mesons are described by the associated effective Lagrangians. According to the heavy quark symmetry, the chiral symmetry, and the hidden local symmetry \cite{Casalbuoni:1992gi,Casalbuoni:1996pg,Yan:1992gz,Bando:1987br,Harada:2003jx}, the effective Lagrangians describing the interactions between the (anti)charmed mesons and the light mesons can be found in  Refs. \cite{Li:2012cs,Ding:2008gr,Sun:2011uh,Wang:2020dya}. 

The key step to obtain the properties of the bound states are to deduce the interaction potentials between the constituents. According to the effective Lagrangians given in Ref. \cite{Li:2012cs}, we can derive the scattering amplitudes, which can be directly related to the effective potentials in momentum space by the Breit approximation \cite{Berestetsky:1982}. The effective potentials in coordinate space can then be obtained by the Fourier transformation \cite{Liu:2009qhy,Wang:2021ajy}. 

In this work, we use the monopole form factor $\mathcal{F}(q^2,m_E^2)=(\Lambda^2-m_E^2)/(\Lambda^2-q^2)$ in the interaction potentials to describe the finite-size effect of the interacting hadrons and manipulate
the off shell effect of the exchanged light mesons, where $q$ and $m_E$ are the momentum and mass of the exchanged meson, respectively. The cutoff $\Lambda$ is a free parameter. According to the experience of studying the deuteron \cite{Tornqvist:1993vu,Tornqvist:1993ng,Chen:2017jjn,Wang:2019nwt}, it should be about 1 GeV, which is a typical scale in low-energy physics. The cutoff depends on the particular application. In this work, we set $\Lambda$ in the range of 0.80$-$2.00 GeV for the $D\bar{D}^*$ system. The $D_s\bar{D}_s^*$ system is harder to bind than the $D\bar{D}^*$ system, so its cutoff is taken in the range of 1.10$-$3.00 GeV.

To show the effective potentials in coordinate space for the $D_s\bar{D}_s^*$ system with the one-boson-exchange model, we have 
\begin{eqnarray}
  V_{\phi}^D(r)&=&-\frac{1}{2}\beta^2g_v^2({\bm\epsilon^{\dagger}_4}\cdot\bm\epsilon_2)Y(\Lambda_i,m_{\phi_i},r),\nonumber\\
  V_{\eta}^C(r)&=&-\frac{2}{9}\frac{g^2}{f_{\pi}^2}\left[({\bm\epsilon^{\dagger}_3}\cdot\bm\epsilon_2)\nabla^2+T({\bm\epsilon^{\dagger}_3},{\bm\epsilon_2})\mathcal{T}\right]Y(\Lambda_i,m_{\eta_i},r),\nonumber\\
  V_{\phi}^C(r)&=&\frac{2}{3}\lambda^2g_v^2\left[({\bm\epsilon^{\dagger}_3}\cdot\bm\epsilon_2)\nabla^2-T({\bm\epsilon^{\dagger}_3},{\bm\epsilon_2})\mathcal{T}\right]Y(\Lambda_i,m_{\phi_i},r),\label{cross} \nonumber\\
\end{eqnarray}
where we have the operator $\mathcal{T}=r\frac{\partial}{\partial r}\frac{1}{r}\frac{\partial}{\partial r}$ and the tensor force operator is defined by $T({\bf A},{\bf B})=3({\bf A}\cdot\hat{r})({\bf B}\cdot\hat{r})-{\bf A}\cdot{\bf B}$. The superscripts $D$ and $C$ indicate the direct and cross channels, respectively. The function of $Y(\Lambda_i,m_i,r)$ is
\begin{eqnarray}
Y(\Lambda_i,m_i,r)&=&\frac{e^{-m_i r}-e^{-\Lambda_i r}}{4\pi r}-\frac{\Lambda^2_i-m^2_i}{8\pi\Lambda_i}e^{-\Lambda_i r},\quad|q_i|\leqslant m\nonumber \\
\end{eqnarray}
with $\Lambda_i=\sqrt{\Lambda^2-q^2_i}$, $m_i=\sqrt{m^2-q^2_i}$, and
$
q_i^2=\frac{\left(m_A^2+m_D^2-m_B^2-m_C^2\right)^2}{\left(2m_C+2m_D\right)^2}.
$
The total effective potentials for the $D_s\bar{D}_s^*$ system can be described as  
\begin{eqnarray}
V_{ D_s\bar{D}^*_s}(r)= V_{\phi}^D(r)+c\left(V_{\eta}^C(r)+V_{\phi}^C(r)\right),
\end{eqnarray}
where $c=-$ and $+$ correspond to the quantum numbers $J^{PC}=1^{++}$ and $1^{+-}$ \cite{Liu:2007bf,Li:2012cs,Wang:2020dya}, respectively. 

The cross diagram of the pion exchange will lead to an infinite in the propagator due to the $m_{D^*}-m_{D}>m_{\pi}$. As a result,  the effective potential in the coordinate space obtained through the Fourier transformation integral contains two parts; the principle value and the imaginary parts. The principle value part corresponds to an oscillatory potential in the coordinate space and has been considered with the static approximation in our present calculations, while the imaginary part arising from the three-body ($D\bar{D}\pi$) cut is very small. This assertion can be verified by some theoretical studies. Specifically, the three-body $D\bar{D}\pi$ decay width of the $X(3872)$ has been calculated as 26 keV \cite{Cheng:2022qcm}, 43$-$56 keV \cite{Schmidt:2018vvl}, and 55$-$65 keV \cite{Baru:2011rs}, respectively, among which the first result arises only from the infinite contribution of one-pion exchange and the other results further consider the dynamical width of $D^*$. It is definite that these values are the order of keV. In addition to the width, this imaginary infinite would also affect the binding energy, and we expect such impact can be partly absorbed into some parameters of the one-boson-exchange model. 


We take Case \Rmnum{3} of the $X(3872)$ as an example to illustrate how we  simultaneously consider the $S$-$D$ wave mixing effect, the coupled channel effect as well as the Coulomb correction \cite{Wang:2019nwt}. The corresponding effective potential $\mathcal{V}$ and kinetic terms $\mathcal{K}$ and $\mathcal{K}_L$ can be expressed as
\begin{widetext}
\begin{equation}
\begin{split}
\setlength{\arraycolsep}{0.5pt}
\mathcal{V}=&
\left(
\begin{matrix}
\mathcal{V}^{D^0\bar{D}^{*0}|^3S_1\rangle\rightarrow D^0\bar{D}^{*0}|^3S_1\rangle}
&\mathcal{V}^{D^0\bar{D}^{*0}|^3S_1\rangle\rightarrow D^0\bar{D}^{*0}|^3D_1\rangle}
&\mathcal{V}^{D^0\bar{D}^{*0}|^3S_1\rangle\rightarrow D^+D^{*-}|^3S_1\rangle}
&\mathcal{V}^{D^0\bar{D}^{*0}|^3S_1\rangle\rightarrow D^+D^{*-}|^3D_1\rangle}\\
\mathcal{V}^{D^0\bar{D}^{*0}|^3D_1\rangle\rightarrow D^0\bar{D}^{*0}|^3S_1\rangle}
&\mathcal{V}^{D^0\bar{D}^{*0}|^3D_1\rangle\rightarrow D^0\bar{D}^{*0}|^3D_1\rangle}
&\mathcal{V}^{D^0\bar{D}^{*0}|^3D_1\rangle\rightarrow D^+D^{*-}|^3S_1\rangle}
&\mathcal{V}^{D^0\bar{D}^{*0}|^3D_1\rangle\rightarrow D^+D^{*-}|^3D_1\rangle}\\
\mathcal{V}^{D^+D^{*-}|^3S_1\rangle\rightarrow D^0\bar{D}^{*0}|^3S_1\rangle}
&\mathcal{V}^{D^+D^{*-}|^3S_1\rangle\rightarrow D^0\bar{D}^{*0}|^3D_1\rangle}
&\mathcal{V}^{D^+D^{*-}|^3S_1\rangle\rightarrow D^+D^{*-}|^3S_1\rangle}
&\mathcal{V}^{D^+D^{*-}|^3S_1\rangle\rightarrow D^+D^{*-}|^3D_1\rangle}\\
\mathcal{V}^{D^+D^{*-}|^3D_1\rangle\rightarrow D^0\bar{D}^{*0}|^3S_1\rangle}
&\mathcal{V}^{D^+D^{*-}|^3D_1\rangle\rightarrow D^0\bar{D}^{*0}|^3D_1\rangle}
&\mathcal{V}^{D^+D^{*-}|^3D_1\rangle\rightarrow D^+D^{*-}|^3S_1\rangle}
&\mathcal{V}^{D^+D^{*-}|^3D_1\rangle\rightarrow D^+D^{*-}|^3D_1\rangle}\\
\end{matrix}
\right),\\
\end{split}
\end{equation}
\end{widetext}
\begin{eqnarray}
\mathcal{K} &=& {\rm diag}\left(-\frac{\nabla^2}{2\mu_1},-\frac{\nabla^2}{2\mu_1},-\frac{\nabla^2}{2\mu_2}+\Delta m,-\frac{\nabla^2}{2\mu_2}+\Delta m\right),\\
\mathcal{K}_L &=& {\rm diag}\left(0,\frac{3}{\mu_1r^2},0,\frac{3}{\mu_2r^2}\right),\label{eq:kinetickl}
\end{eqnarray}
where $\Delta m=m_{D^+}+m_{D^{*-}}-m_{D^0}-m_{\bar{D}^{*0}}$, and $\mu_1=\frac{m_{D^0}m_{\bar{D}^{*0}}}{m_{D^0}+m_{\bar{D}^{*0}}}$ and $\mu_2=\frac{m_{D^+}m_{D^{*-}}}{m_{D^+}+m_{D^{*-}}}$ are the reduced masses of the $D^0\bar{D}^{*0}$ and $D^+D^{*-}$ systems, respectively. Specially, when we consider the $S$-$D$ wave mixing effect, there are the centrifugal potentials $\mathcal{K}_L$ in Eq.~(\ref{eq:kinetickl}). In addition, we add the Coulomb interaction to the third and fourth diagonal elements of the potential term, respectively. With the above preparations, we can obtain the corresponding bound state solutions, including the binding energy $E$, the root-mean-square radius $r_{\rm RMS}$ and the probabilities of different components, by solving the coupled channel Schr\"{o}dinger equation
\begin{eqnarray}\label{Schrodinger}
\left(\mathcal{K}+\mathcal{K}_L+\mathcal{V}\right)\hat\phi(r)=E\hat\phi(r)
\end{eqnarray}
with
\begin{equation}
\hat\phi(r)=(\phi_{[D^0\bar{D}^{*0}]|^3S_1\rangle},\,\phi_{[D^0\bar{D}^{*0}]|^3D_1\rangle},\,\phi_{[D^+D^{*-}]|^3S_1\rangle},\,\phi_{[D^+D^{*-}]|^3D_1\rangle})^T.
\end{equation}

\section{Numerical results for formation with Coulomb interaction}\label{sec3}
In this section, we show the numerical results for the properties of the $D\bar{D}^*$ ($1^{++}$), $D\bar{D}^*$ ($1^{+-}$), $D_s\bar{D}_s^*$ ($1^{++}$), and $D_s\bar{D}_s^*$ ($1^{+-}$) systems, and discuss the role of the Coulomb interaction in the formation. We use the FORTRAN program FESSDE \cite{Abrashkevich:1995tb,Abrashkevich:1998nm} to solve the coupled channel Schr\"{o}dinger equation and get the binding energy \textit{E}, the  root-mean-square radius $r_{{\rm RMS}}$ and the probabilities of the different components.

First we need to determine the parameters in our model. The couplings associated with the vector meson exchange are $g_v=5.8$ and $\beta=0.9$, which are obtained by the vector meson dominance mechanism \cite{Bando:1987br,Isola:2003fh}. By comparing the form factor, the coupling $\lambda$ is determined to be $0.56~\mbox{GeV}^{-1}$ \cite{Casalbuoni:1996pg,Isola:2003fh}. The scalar meson coupling is $g_s=g_{\pi}/(2\sqrt{6})$ with $g_\pi=3.73$ \cite{Liu:2008xz}. The meson masses are taken from the PDG \cite{ParticleDataGroup:2022pth}.

\subsection{The bound state solutions of the $X(3872)$ state}

The $X(3872)$ has the small binding energy (less than 1~MeV) and narrow width, which implies that it may be a loosely bound state composed of $D\bar{D}^{*}$. We take the cutoff parameter $\Lambda$ in the range of 0.80$-$2.00 GeV to investigate the binding properties of the $X(3872)$, and the results are presented in Table \ref{X(3872)}. Even with uncertainties, the binding of the $X(3872)$ is now fairly well understood. We discuss the bound state properties of the $X(3872)$ in the range, where its binding energy is less than 1 MeV.
\begin{table*}[htbp]
\caption{The bound-state solutions of the $X(3872)$ as the possible $D\bar{D}^*$ molecular state with different form factors like $\mathcal{F}(q^2,m_E^2)=(\Lambda^2-m_E^2)/(\Lambda^2-q^2)$ and $\mathcal{F}(q^2)=\Lambda^2/(\Lambda^2-q^2)$ for the Cases \Rmnum{1}$-$\Rmnum{3}. The ``\ding{55}" means that the binding solution does not exist and the Coulomb interaction is $-\alpha_{\rm EM}/r$ in Case \Rmnum{3}. The $P_i$ means the probability of the $i$th channel as written in Table \ref{Channel}.}\label{X(3872)}
\renewcommand\arraystretch{1.35}
\centering
\setlength{\tabcolsep}{0.6mm}
\begin{tabular*}{180mm}{@{\extracolsep{\fill}}c
cccc
cccccc
cccccc}
\toprule[1pt]
\toprule[1pt]
\multicolumn{17}{c}{$\mathcal{F}(q^2,m_E^2)=(\Lambda^2-m_E^2)/(\Lambda^2-q^2)$}\\
\toprule[0.75pt]
\multirow{2}*{$\Lambda$(GeV)}&\multicolumn{4}{c}{Case \Rmnum{1}}&\multicolumn{6}{c}{Case \Rmnum{2}}&\multicolumn{6}{c}{Case \Rmnum{3}}\\
\Xcline{2-5}{0.75pt}
\Xcline{6-11}{0.75pt}
\Xcline{12-17}{0.75pt}

                            ~&\textit{E}(MeV)&$r_{\rm RMS}$(fm)&$P_1$&$P_2$
                            &\textit{E}(MeV)&$r_{\rm RMS}$(fm)&$P_1$&$P_2$&$P_3$&$P_4$
                            &\textit{E}(MeV)&$r_{\rm RMS}$(fm)&$P_1$&$P_2$&$P_3$&$P_4$\\

\midrule[0.75pt]
1.08 &$-0.12$&5.84&99.16&0.84  &\ding{55}&~&~&~&~&~                    &\ding{55}&~&~&~&~&~\\
1.16 &$-2.13$&2.51&98.24&1.76&\ding{55}&~&~&~&~&~                                      &$-0.25$&4.90&86.82&0.49&12.19&0.50\\
1.17 &$-2.54$&2.32&98.14&1.86&$-0.25$&4.94&88.02&0.48&11.01&0.49                      &$-0.50$&4.06&83.17&0.61&15.61&0.61\\
1.18 &$-2.98$&2.16&98.04&1.96&$-0.49$&4.12&84.67&0.60&14.13&0.60                      &$-0.82$&3.39&79.68&0.71&18.89&0.72\\
1.19 &$-3.46$&2.03&97.95&2.05&$-0.80$&3.44&81.40&0.70&17.19&0.71                      &$-1.19$&2.89&76.51&0.81&21.86&0.81\\
\toprule[0.75pt]
\multicolumn{17}{c}{$\mathcal{F}(q^2)=\Lambda^2/(\Lambda^2-q^2)$}\\
\toprule[0.75pt]
\multirow{2}*{$\Lambda$(GeV)}&\multicolumn{4}{c}{Case \Rmnum{1}}&\multicolumn{6}{c}{Case \Rmnum{2}}&\multicolumn{6}{c}{Case \Rmnum{3}}\\
\Xcline{2-5}{0.75pt}
\Xcline{6-11}{0.75pt}
\Xcline{12-17}{0.75pt}

                            ~&\textit{E}(MeV)&$r_{\rm RMS}$(fm)&$P_1$&$P_2$
                            &\textit{E}(MeV)&$r_{\rm RMS}$(fm)&$P_1$&$P_2$&$P_3$&$P_4$
                            &\textit{E}(MeV)&$r_{\rm RMS}$(fm)&$P_1$&$P_2$&$P_3$&$P_4$\\
\midrule[0.75pt]
0.40 &$-0.13$&6.08&99.60&0.40  &\ding{55}&~&~&~&~&~                        &\ding{55}&~&~&~&~&~\\
0.54 &$-1.91$&2.86&99.28&0.72&\ding{55}&~&~&~&~&~                                                        &$-0.18$&5.46&90.72&0.20&8.91&0.17\\
0.55 &$-2.13$&2.73&99.25&0.75  &$-0.16$&5.56&91.80&0.20&7.84&0.17                    &$-0.29$&4.99&89.06&0.22&10.52&0.19\\
0.56 &$-2.36$&2.61&99.22&0.78  &$-0.27$&5.10&90.32&0.22&9.27&0.19                    &$-0.42$&4.54&87.30&0.25&12.22&0.22\\
0.59 &$-3.13$&2.32&99.12&0.88 &$-0.70$&3.86&85.41&0.31&14.01&0.27                    &$-0.93$&3.43&81.88&0.34&17.48&0.30\\

\bottomrule[1pt]
\bottomrule[1pt]
\end{tabular*}
\end{table*}

In Table \ref{X(3872)}, bound state solutions in Case \Rmnum{1} are given when the cutoff parameter is 1.08 GeV, where the corresponding binding energy and root-mean-square radius are $-0.12$ MeV and 5.84 fm, respectively. After taking into account the isospin breaking effect due to the hadron mass difference, we can obtain a loosely bound state with a binding energy of $-0.25$ MeV and a  root-mean-square radius of 4.94 fm for a cutoff parameter of 1.17 GeV in Case \Rmnum{2}. Clearly, the isospin breaking effect weakens the attraction between the composed hadrons in the $X(3872)$. If we add the Coulomb interaction $-\alpha_{\rm EM}/r$, the binding energy increases to $-0.50$ MeV and the root-mean-square radius decreases to 4.06 fm for the cutoff 1.17 GeV in Case \Rmnum{3}. Compared to Case \Rmnum{2}, the binding energy of Case \Rmnum{3} increases by about 0.2 MeV to 0.4 MeV with the same cutoff parameter after adding the Coulomb interaction as shown in Table \ref{X(3872)}. This variation  in the binding energy caused by the Coulomb interaction is not small compared to the tiny binding energy of the $X(3872)$.
\begin{figure}[tbp]
\includegraphics[width=250pt]{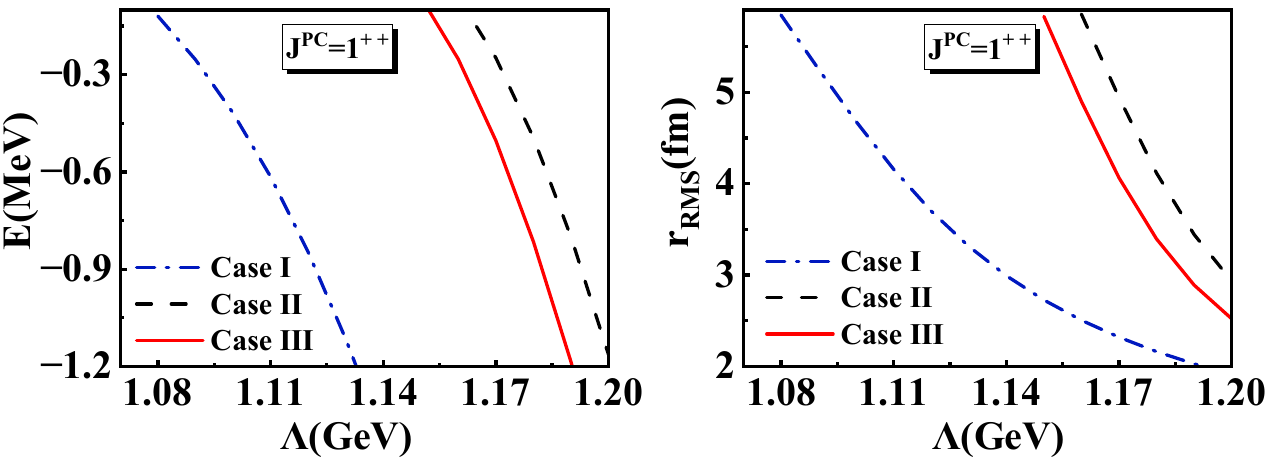}
\caption{Comparison for the binding energy and root-mean-square radius of the $X(3872)$ among three cases. See the caption of Table \ref{Channel} for detailed meanings of Case \Rmnum{1}$-$Case \Rmnum{3}. 
}\label{Er1++}
\end{figure}

For the exact $D\bar{D}^*$ isospin singlet, the dominant channel is $[D\bar{D}^{*}]|^3S_1\rangle$, with a probability of 99.16\% and the $D$-wave channel with 0.84\% when the cutoff is $\Lambda=1.08$ GeV. For Case \Rmnum{2}, $S$-wave channels also play the leading role with the probability of $(88.02\%+11.01\%)=99.03\%$ after considering isospin breaking. For the same binding energy $-0.25$ MeV, the probabilities ($P_3$, $P_4$) for the charged channels increase after adding the Coulomb interaction (see the comparison of the results for Case \Rmnum{2} and Case \Rmnum{3}). 

To clearly show the Coulomb interaction on the binding properties of the $X(3872)$, in Fig. \ref{Er1++}, we plot the binding energies $E$ and the the root-mean-square radii $r_{\rm RMS}$ corresponding to three different cases. The effect of the Coulomb interaction is smaller compared to the contribution of the isospin breaking due to the meson mass difference. The Coulomb interaction makes the binding energy $|E|$ larger and the the root-mean-square radius $r_{\rm RMS}$ smaller as shown by the curves 
corresponding to Cases \Rmnum{2} and \Rmnum{3} in Fig. \ref{Er1++}.

For the monopole form factor $\mathcal{F}(q^2,m_E^2)=(\Lambda^2-m_E^2)/(\Lambda^2-q^2)$, the factor $(\Lambda^2-m_E^2)$ may reduce the strength of the vector meson exchange interactions. In order to further test our conclusions, we use a different kind of form factor $\mathcal{F}(q^2)=\Lambda^2/(\Lambda^2-q^2)$ to recalculate the bound solutions of the $X(3872)$ in Table \ref{X(3872)}. Comparing the numerical results between two form factors, we would like to mention that the cutoff values around 1.0 GeV in the form factor $\mathcal{F}(q^2,m_E^2)=(\Lambda^2-m_E^2)/(\Lambda^2-q^2)$ and 0.5 GeV in the form factor $\mathcal{F}(q^2)=\Lambda^2/(\Lambda^2-q^2)$ are the reasonable input parameters to study the possible hadronic molecular states \cite{Chen:2017vai,Wang:2021ajy}. We can obtain the bound state solution by setting the cutoff values around 1.16 GeV in the form factor $\mathcal{F}(q^2,m_E^2)$ and 0.55 GeV in the form factor $\mathcal{F}(q^2)$ for the Case \Rmnum{3}. We can see that the form factor does not have much influence on the binding energy and the probabilities of the different components, reflecting that the wave function is stable with two types of form factors. It will be advantageous for us to discuss the nature of the $X(3872)$ later.

\begin{table*}[hbtp]
\caption{The bound state solutions of the \textit{C}-parity partner of $D\bar{D}^*$ with $J^{PC}=1^{+-}$ for Cases \Rmnum{1}$-$\Rmnum{3}. The ``\ding{55}" means the binding solution does not exist and the Coulomb interaction is $-\alpha_{\rm EM}/r$ in Case \Rmnum{3}. The $P_i$ means the probability of the $i$th channel as written in Table \ref{Channel}. }\label{partner state}
\renewcommand\arraystretch{1.25}
\centering
\setlength{\tabcolsep}{0.6mm}
\begin{tabular*}{180mm}{@{\extracolsep{\fill}}c
cccccc
cccccc
cccccc}
\toprule[1pt]
\toprule[1pt]
\multirow{3}*{\makecell[c]{$\Lambda$ (GeV)}}&\multicolumn{4}{c}{Case \Rmnum{1}}&\multicolumn{6}{c}{Case \Rmnum{2}}&\multicolumn{6}{c}{Case \Rmnum{3}}\\
\Xcline{2-5}{0.75pt}
\Xcline{6-11}{0.75pt}
\Xcline{12-17}{0.75pt}

                     ~&\textit{E} (MeV)&$r_{\rm RMS}$ (fm)&$P_1$&$P_2$&\textit{E} (MeV)&$r_{\rm RMS}$ (fm)&$P_1$&$P_2$&$P_3$&$P_4$
                            &\textit{E} (MeV)&$r_{\rm RMS}$ (fm)&$P_1$&$P_2$&$P_3$&$P_4$\\
~&~&~&$P_3$&$P_4$&~&~&$P_5$&$P_6$&$P_7$&$P_8$&~&~&$P_5$&$P_6$&$P_7$&$P_8$\\                            
\midrule[1pt]                            
1.16&$-0.20$&5.42&96.29&2.19          &\ding{55}&~&~&~&~&~&\ding{55}&~&~&~&~&~&\\
~&~&~&1.06&0.46\\
\addlinespace[0.55em]
1.19&$-1.92$&2.57&91.14&4.50
&\ding{55}&~&~&~&~&~&$-0.14$&5.35&87.16&1.12&8.46&1.09\\
~&~&~&3.08&1.28&         ~&~&~&~&~&~
~&~&~&0.72&0.30&0.81&0.33\\
\addlinespace[0.55em]
1.20&$-2.88$&2.15&89.44&5.15
&$-0.42$&4.27&83.32&1.49&10.71&1.14&$-0.73$&3.48&77.56&1.84&14.97&1.78\\
~&~&~&3.84&1.58&         ~&~&1.02&0.43&1.12&0.46
~&~&~&1.31&0.54&1.42&0.58\\
\addlinespace[0.55em]
1.23&$-6.94$&1.48&84.61&6.75
&$-3.60$&1.72&61.96&3.10&24.27&2.96&$-4.30$&1.59&57.99&3.24&27.41&3.11\\
~&~&~&6.17&2.47&         ~&~&2.70&1.09&2.79&1.13
~&~&~&2.90&1.16&3.00&1.19\\
\addlinespace[0.55em]

1.25&$-10.63$&1.25&81.59&7.62        &$-7.04$&1.31&54.13&3.72&28.36&3.55
&$-7.91$&1.26&51.16&3.79&30.73&3.65\\
~&~&~&7.72&3.07&~&~&3.64&1.45&3.68&1.48
&~&~&3.79&1.50&3.86&1.52\\
\bottomrule[1pt]
\bottomrule[1pt]
\end{tabular*}
\end{table*} 
\subsection{The bound solutions of the $X(3872)$ partners}

In this section, we present the numerical results for the partner states of the $X(3872)$ with the Coulomb interaction. In detail, we discuss the systems of $D\bar{D}^*$ with $J^{PC}=1^{+-}$, $D_s\bar{D}_s^*$ with $J^{PC}=1^{++}$, and $D_s\bar{D}_s^*$ with $J^{PC}=1^{+-}$.

\begin{figure}[btp]
\includegraphics[width=250pt]{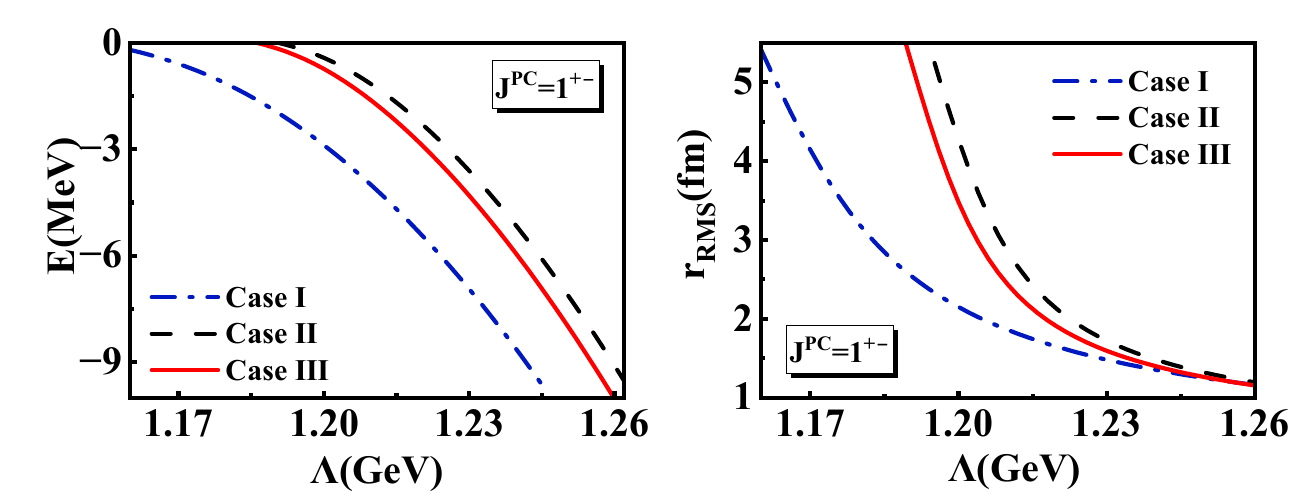}
\caption{Comparison of the binding energy and root-mean-square radius for the $D\bar{D}^*$ system with $J^{PC}=1^{+-}$ among three cases. See the caption of Table \ref{Channel} for detailed meaning of Case \Rmnum{1}$-$Case \Rmnum{3}. 
}\label{Er1+-}
\end{figure}

For the odd $C$-parity $D\bar{D}^*$ system with $J^{PC}=1^{+-}$, we introduce the couplings of $[D\bar{D}^*$]($|^3S_1\rangle$, $|^3D_1\rangle$) to $[D^*\bar{D}^*]$($|^3S_1\rangle$, $|^3D_1\rangle$). The numerical results of the $[D\bar{D}^*]$ system with $J^{PC}=1^{+-}$ for the three cases are presented in Table \ref{partner state}. We can obtain a loosely bound solution with binding energy $-0.20$ MeV and root-mean-square radius 5.42 fm when cutoff is fixed at 1.16 GeV in Case \Rmnum{1}. We also take into account the explicit mass splitting between the charged and neutral $D$ ($D^*$) mesons in Case \Rmnum{2}. The required cutoff parameter in Case \Rmnum{2} is larger than that in Case \Rmnum{1} and we can find bound state solutions when cutoff parameter is fixed at 1.20 GeV. In other words, the isospin breaking effect weakens the attraction. We further consider Coulomb interaction that is from both charged meson pair $D^+D^{*-}$ and $D^{*+}D^{*-}$ pair in Case \Rmnum{3}. Compared with the binding solutions of Case \Rmnum {2} with the cutoff 1.20 GeV, the binding energy $|E|$ increases by $0.31$ MeV and root-mean-square radius decreases by 0.79 fm in Case \Rmnum{3}. This indicates that the Coulomb contribution from four charged channels ($[D^+D^{*-}]$$|^3S_1\rangle$, $[D^+D^{*-}]$$|^3D_1\rangle$, $[D^{*+} D^{*-}]$$|^3S_1\rangle$, $[D^{*+}D^{*-}]$$|^3D_1\rangle$) 
strengthens the interaction between the components. Similarly, the dominant channel is still $S$-wave $[D^0\bar{D}^{*0}]$ and $[D^+D^{*-}]$ with the probability of 92.53\% for Case \Rmnum{3}, if the cutoff is fixed at 1.20 GeV. In order to make clear the role of the Coulomb interaction in the formation of the bound state for the partner state $[D\bar{D}^*]$ with $J^{PC}=1^{+-}$, we compare the bound solutions corresponding to three cases in Fig. \ref{Er1+-}. The relation of binding energy and root-mean-square radius can be obtained by comparing three cases with the same cutoff, i.e., $|E|^{\rm Case\, \rm{\Rmnum{1}}}$\textgreater $ |E|^{\rm{Case\, \Rmnum{3}}}$\textgreater$ |E|^{\rm{Case\, \Rmnum{2}}}$ and $r_{\rm RMS}^{\rm{Case\, \Rmnum{2}}}$\textgreater$ r_{\rm RMS}^{\rm{Case\, \Rmnum{3}}}$\textgreater$ r_{\rm RMS}^{\rm{Case\, \Rmnum{1}}}$.

For the $D_s\bar{D}_s^*$ system, the charmed-strange meson $D_s$ constitutes an isospin singlet, i.e., there is only the exclusive charged mode in the $D_s\bar{D}_s^*$ system. It is advantageous for us to study the role of the Coulomb interaction in the formation of the bound states of hadronic molecular. We obtain the bound state solutions taking into account  the $S$-$D$ wave mixing effect. We vary the cutoff values of $\Lambda$ from 1.10 GeV to 3.00 GeV. The numerical results can be found in Table \ref{DsDs*}. We can see that the $S$-wave contribution dominates since $P_S \gg P_D$.
\setlength{\tabcolsep}{2.2mm}
\begin{table*}[hbtp]
\caption{The bound state solutions of the $D_s\bar{D}^*_s$ system with $J^{PC}=1^{++}$ and $1^{+-}$ for the cases with and without Coulomb interaction. The $S$-$D$ mixing effect is taken into account in both cases. The ``\ding{55}" means that the binding solution does not exist. The $P_S$ means that the probability of the $S$-wave $[D_s\bar{D}^*_s]$ and the probability of the $D$-wave $[D_s\bar{D}^*_s]$ can be obtained by $P_D=1-P_S$. }\label{DsDs*}
\renewcommand\arraystretch{1.35}
\centering
\setlength{\tabcolsep}{0.75mm}
\begin{tabular*}{180mm}{@{\extracolsep{\fill}}cc
ccc
ccc}

\toprule[1pt]
\toprule[1pt]
\multirow{2}{*}{$J^{PC}$}&\multirow{2}{*}{$\Lambda$}&\multicolumn{3}{c}{\small No Coulomb Interaction}&\multicolumn{3}{c}{\small With Coulomb Interaction}\\
\Xcline{3-5}{0.75pt}
\Xcline{6-8}{0.75pt}

 ~&(GeV)&\textit{E}(MeV)&$r_{\rm RMS}$(fm)&$P_S$&\textit{E}(MeV)&$r_{\rm RMS}$(fm)&$P_S$\\
\midrule[0.75pt]
$1^{++}$ &2.50     &\ding{55}&~&~         &$-0.15$&6.61&99.32  \\
~&2.69     &$-0.14$&5.56&99.56   &$-1.44$&3.05&98.88\\
                       ~&2.75     &$-0.61$&3.92&99.23   &$-2.28$&2.42&98.59\\
                       ~&2.85     &$-2.01$&2.36&98.62   &$-4.26$&1.78&98.06\\

$1^{+-}$&2.21  &\ding{55}&~&~          &$-0.26$ &5.97&99.45\\
                      ~&2.26  &$-0.23$&5.10&99.41    &$-1.12$&3.41&99.16\\
                      ~&2.30  &$-1.34$&2.79&98.73    &$-2.59$&2.21&98.64\\
                      ~&2.40  &$-8.41$&1.19&97.03    &$-10.32$&1.12&97.12\\
\bottomrule[1pt]
\bottomrule[1pt]
\end{tabular*}
\end{table*}

For the $D_s\bar{D}_s^*$ system with $J^{PC}=1^{++}$ state, we obtain a loosely bound state with a binding energy of $-0.14$ MeV for the cutoff being of 2.69 GeV and corresponding root-mean-square radius is 5.56 fm without Coulomb interaction. As shown in Table \ref{DsDs*}, the cutoff required for the bound solution is less than 2.69 GeV after adding the Coulomb interaction. It is obvious that the negative $C$-parity $D_s\bar{D}_s^*$ is easier to bind than the positive one from Table \ref{DsDs*}. The binding energy increases by about 1.3$-$2.2 MeV and the root-mean-square radius decreases by about 0.5$-$2.5 fm due to the Coulomb interaction. The Coulomb interaction makes the binding energy bigger.

\begin{figure}[tbp]
\includegraphics[width=250pt]{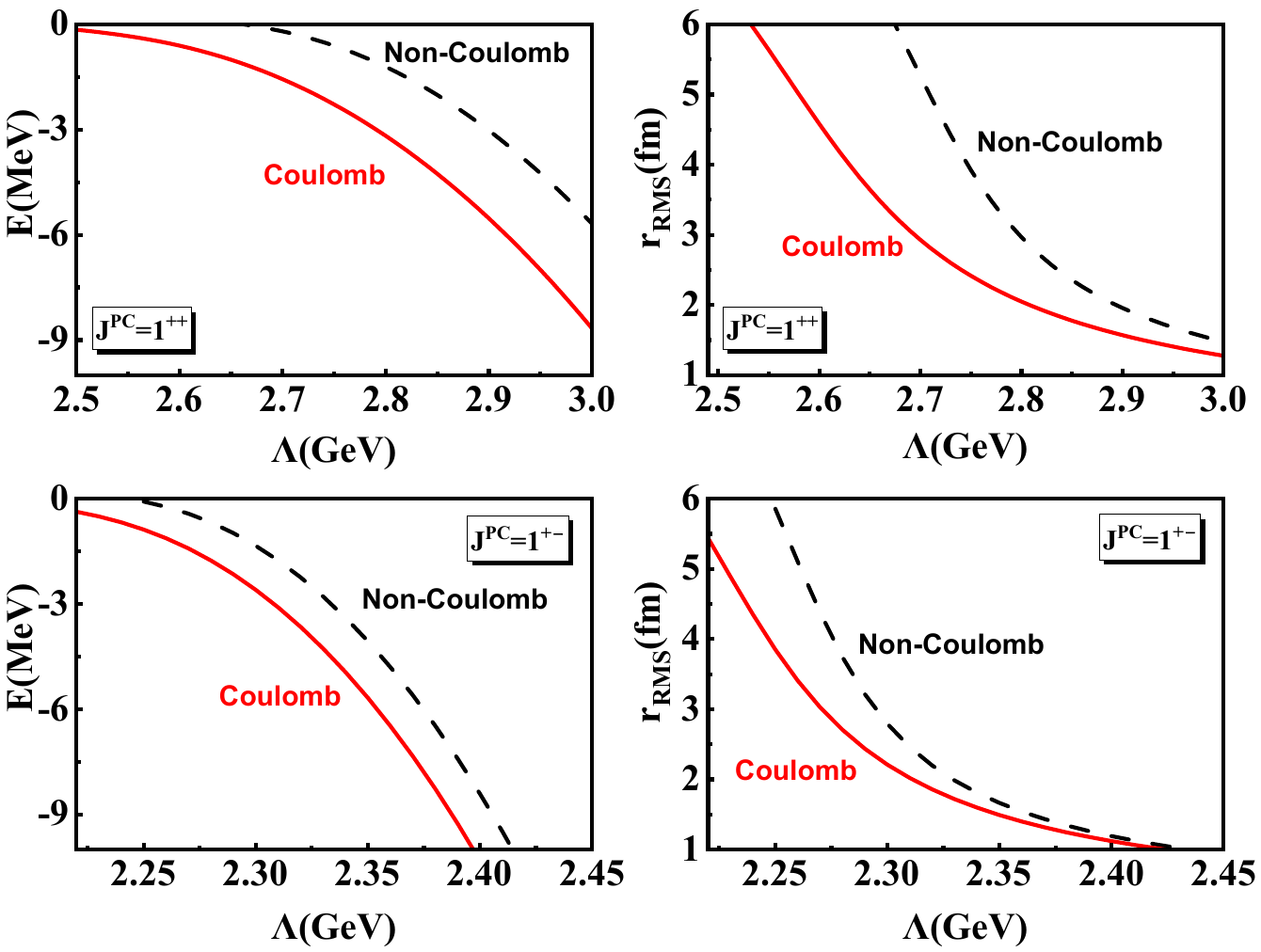}
\caption{Comparison of the binding energy and root-mean-square radius of the $D_s\bar{D}_s^*$ system for the cases with and without Coulomb interaction. 
}\label{DsDs}
\end{figure}

Similar to the $D\bar{D}^*$ system, we compare the binding energies $E$ and root-mean-square radii $r_{\rm RMS}$ of the $1^{++}$ and $1^{+-}$ $D_s\bar{D}_s^*$ systems with and without Coulomb interaction in Fig.~\ref{DsDs}. The Coulomb interaction causes the binding energy $E$ and root-mean-square radius $r_{\rm RMS}$ to deviate significantly from the figure. Since the $D_s\bar{D}_s^*$ with $J^{PC}=1^{+-}$ has a smaller cutoff that drops more contributions from large momentum charmed-strange mesons, it would be more likely to form a loosely bound state than the $1^{++}$ state.

The reason why the Coulomb effect is more manifest for the $D_s\bar{D}_s^*$ system is that the strong interaction between $D_s$ and $\bar{D}_s^*$ is less than $D\bar{D}^*$. The $\eta$ and $\phi$ exchange interaction is much smaller than the long-range pion exchange interaction. The pion exchange interaction is usually dominant in the formation of the loosely bound states with the one-boson-exchange model \cite{Tornqvist:1993ng}. However, the one pion exchange is forbidden for the $D_s\bar{D}_s^*$ system due to parity conservation. So the $D\bar{D}^*$ system would be bound deeper. Another reason is that all the $D_s^{(*)}$ mesons are charged while the neutral channels are dominant in the $D\bar{D}^*$ system, leading to a larger effect of the Coulomb interactions on the binding energy of the $D_s\bar{D}_s^*$ system.

\subsection{The effect of charge distribution on bound state properties}

In general, $D\bar{D}^*$ and its partner states can be regarded as loosely bound states, and the charged $D^{(*)}$ mesons as point particles. Thus the Coulomb interaction between charged particles can be  simply described as $\alpha_{\rm EM}\frac{Q_1Q_2}{r}$, where $Q_1$ and $Q_2$ are the charges carried by the charged particles, and $\alpha_{\rm EM}=1/137$ is dimensionless electromagnetic fine structure constant. However, the charged mesons have sizes and there are the charge distributions in them. To estimate the effect of the charge distribution, we use an exponential charge distribution density \cite{Lyu:2021qsh,Wiringa:1994wb},
\begin{equation}
\rho(r)=\frac{12\sqrt{6}}{\pi r_{d}^3}\exp(-\frac{2\sqrt{6}r}{r_d}),
\end{equation}
where $r_d$ stands for the charge radius. Thus the improved Coulomb interaction can be expressed as \cite{Lyu:2021qsh},
\begin{eqnarray}
V_{\rm Coul.}(r)&=&\alpha_{\rm EM}\iint \rm d^3 \bm{{r}_1} d^3 \bm{{r}_2}\frac{\rho(\bf{r}_1)\rho(\lvert\bf{r}_2-\bf{r} \rvert)}{\lvert \bf{r}_1-\bf{r}_2 \rvert}\nonumber\\
&=&\frac{\alpha_{\rm EM}}{r}F(x),
\end{eqnarray}
where $x=2\sqrt{6}r/r_d$ and $F(x)=1-e^{-x}[1+\frac{11}{16}x+\frac{3}{16}x^2+\frac{1}{48}x^3]$. The charge radius $r_d=\sqrt{|\langle r^2 \rangle_{\rm charge}|}$ of the \textit{D} meson is about 0.39 fm~\cite{Can:2012tx}.

\setlength{\tabcolsep}{2.2mm}
\begin{table}[tbp]
\caption{The bound-state solutions of the $X(3872)$ when we consider the isospin symmetry breaking and the charge distribution of the charmed mesons. The $P_i$ means the probability of the $i$th channel as written in Table \ref{Channel}.}\label{OBE charge distribution}
\renewcommand\arraystretch{1.35}
\begin{tabular*}{85mm}{@{\extracolsep{\fill}}ccccccccccc}
\toprule[1pt]
\toprule[1pt]
$\Lambda$(GeV)&\textit{E}(MeV)&$r_{\rm RMS}$(fm)&$P_1$&$P_2$&$P_3$&$P_4$&\\
\midrule[0.75pt]
1.16&$-0.25$&4.90&86.84&0.49&12.17&0.50\\
1.17&$-0.50$&4.07&83.20&0.60&15.59&0.61\\
1.18&$-0.81$&3.40&79.70&0.71&18.87&0.72\\
1.19&$-1.18$&2.89&76.53&0.81&21.85&0.81\\
\bottomrule[1pt]
\bottomrule[1pt]
\end{tabular*}
\end{table}

We present the binding solutions of the $X(3872)$ with the above Coulomb potential in Table \ref{OBE charge distribution}. Comparing the results in Table~\ref{OBE charge distribution} and Case \Rmnum{3} in Table~\ref{X(3872)}, the difference in binding energy is 0.01 MeV only for $\Lambda=1.18$ GeV. This suggests that the modified Coulomb potential $-\frac{\alpha_{\rm EM}}{r}F(x)$ has almost no effect on the bound state solutions for the $X(3872)$. The reason for this is that the meson charge radius is small (0.39 fm) compared to the root-mean-square radius of 3.40 fm with $\Lambda=1.18$ GeV. It is reasonable to assume that the inner mesons in a loosely bound state $X(3872)$ are point particles. 

\section{radiative decay}\label{sec4}

In this section, we use the effective Lagrangian approach to study the radiative processes of  $X(3872)\rightarrow\gamma \psi_n$ and $[D_{(s)}\bar{D}^*_{(s) } (1^{+-})]\rightarrow\gamma \eta_c$ with $\psi_n=J/\psi$, $\psi(2S)$ and $\eta_c=\eta_c(1S)$, $\eta_c(2S)$ in the pure molecular state framework. Assuming that a bound state formed by two particles $AB$ can be written as $[AB]$, the decay amplitude can be calculated from the scattering process with \cite{Luo:2023hnp,Zhang:2006ix}
\begin{eqnarray}\label{amplitude}
\mathcal{M}^{JM}_{[AB]\rightarrow CD}&=&\frac{\sqrt{2m_{[AB]}}}{\sqrt{2m_A}\sqrt{2m_B}}\int\frac{\rm d^3\textbf{p}}{(2\pi)^{3/2}}\,\hat\phi^{JM}_{[AB]}(\textbf{p})\otimes\hat{\mathcal{M}}_{AB\rightarrow CD}. \nonumber\\
\end{eqnarray}
Here, $\hat\phi^{JM}_{[AB]}(\textbf{p})$ is the wave function of the bound state $[AB]$ in momentum space, which can be expressed as
\begin{eqnarray}\label{phiJM}
\hat\phi^{JM}_{[AB]}(\textbf{p})&=&\bigg\{\phi_{[AB]|^3L_J\rangle}(|\textbf{p}|)\,\,C^{S,m_S}_{1m_1,1m_2}C^{J,M}_{Sm_S,Lm_L}\,Y_{L,m_L}(\theta,\phi) \nonumber\\ 
&&\qquad \bigg|\forall ~ L,S,m_L,m_1,m_2\bigg\}.
\end{eqnarray}
$\hat{\mathcal{M}}_{AB\rightarrow CD}$ is the scattering amplitude of the $AB\rightarrow CD$ process with the quantum number corresponding to the term in Eq. (\ref{phiJM}). $\textbf{p}$ and $\textbf{k}$ are the momenta of $B$ and $C$, respectively. Here, we denote the direction of the momentum $\textbf{p}$ as $(\theta,\phi)$. Taking the $X(3872)\rightarrow \gamma \psi_n$ as an example, the momentum of the final state ($\gamma$ or $\psi_n$) in the center-of-mass frame is $|\textbf{k}|=\frac{m^2_{X}-m^2_{\psi_n}}{2m_X}$.  According to the introduction of the Sec. \ref{sec2}, the wave function of the different components can be obtained by solving the coupled channel Schr\"{o}dinger equation.

Then the decay width is given by
\begin{equation}\label{width1}
\Gamma_{[AB]\rightarrow  CD}=\frac{1}{3}\frac{|\textbf{k}|}{32\pi^2m_{X}^2}\sum_{M}\int \left|\mathcal{M}^{JM}_{[AB]\rightarrow CD}\right|^2{\rm d}\Omega_{\textbf{k}}.
\end{equation}
The factor of $1/{3}$ is due to spin average over the initial state.

\subsection{Feynman diagrams and effective Lagrangians}

\begin{figure}[tbp]
\centering
\includegraphics[width=250pt]{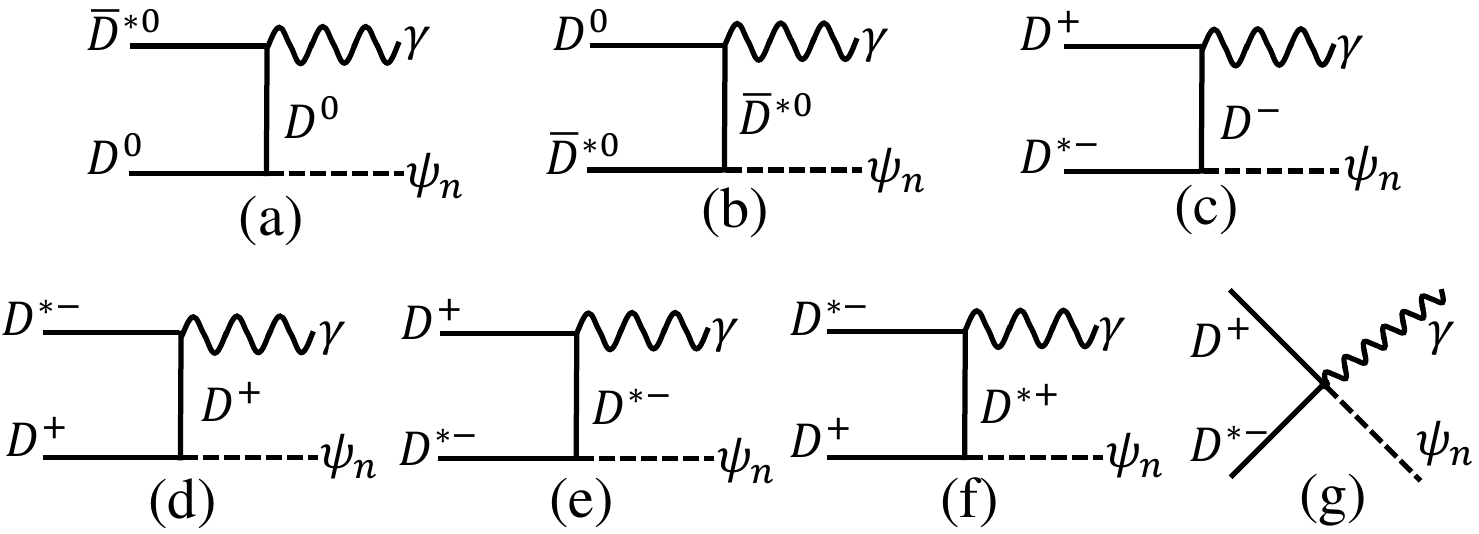}
\caption{The Feynman diagrams for the transitions $X(3872)\rightarrow\gamma\psi_n$ when the $X(3872)$ is a $D\bar{D}^*$ hadronic molecule with $J^{PC}=1^{++}$. The conjugated diagrams are not shown, but are included in the calculations.}\label{tree}
\end{figure}

\begin{figure}[tbp]
\centering
\includegraphics[width=245pt]{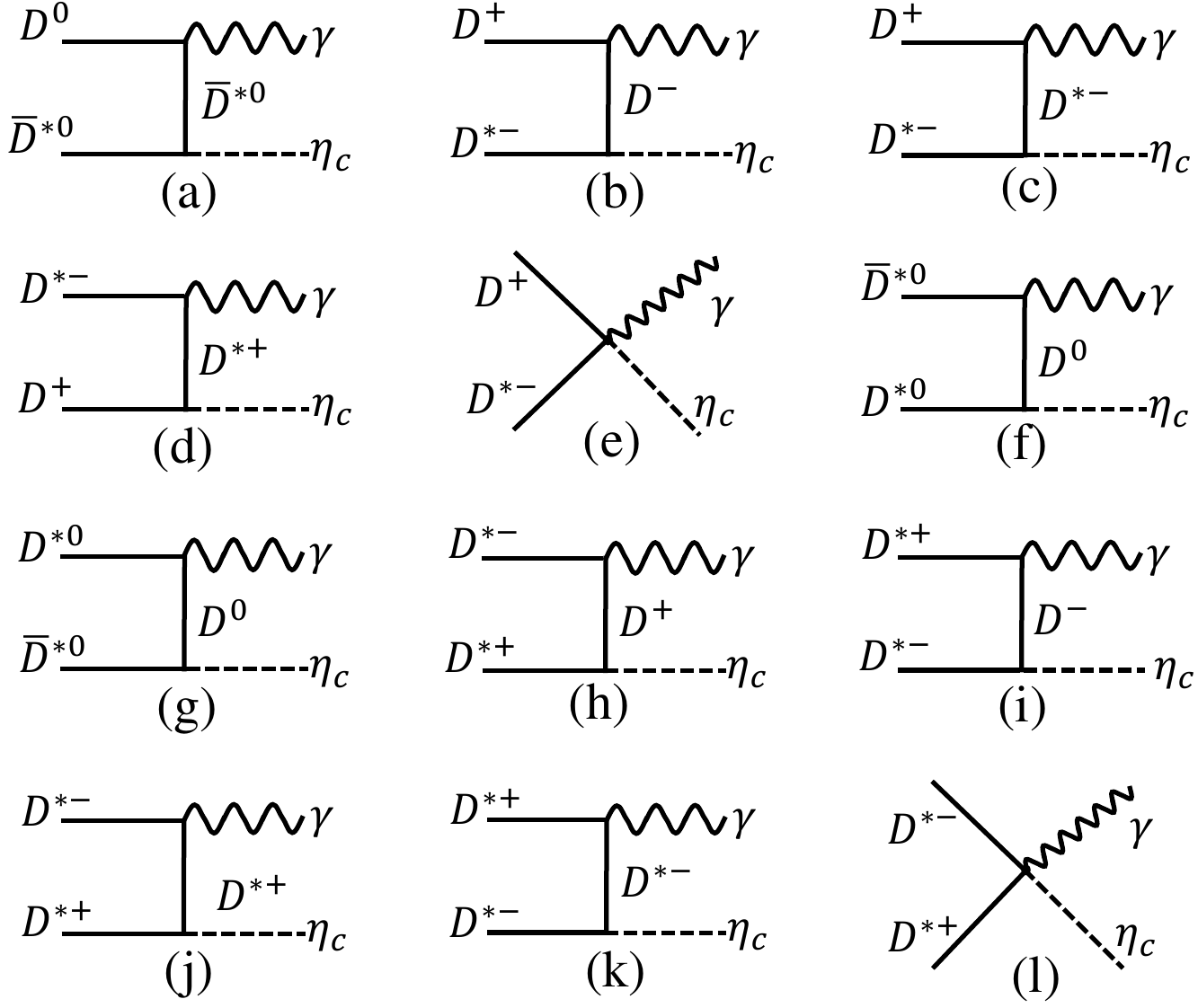}
\caption{The schematic diagrams for $D^*\bar{D}^*$ with $J^{PC}=1^{+-}$ decaying to $\gamma\eta_c$. The conjugated diagrams are not shown, but included in the calculations.}\label{partnerfig}
\end{figure}

We display the Feynman diagrams and the effective Lagrangian in this section. For the $X(3872)$, the diagrams depicting the scattering to $\gamma \psi_n$ are shown in Fig. \ref{tree}.  The contact interaction is required for the scattering process of $D\bar{D}^{*}\rightarrow\gamma \psi_{n}$ to maintain the gauge invariance of the photon fields, as shown in Fig. \ref{tree}~(g). The diagrams describing the process of $D\bar{D}^*$ with $J^{PC}=1^{+-}$ decaying to $\gamma\eta_c$ are shown in Fig. \ref{partnerfig}. 
  
We list the corresponding interaction Lagrangians of these processes in Eqs. (\ref{gamma})$-$(\ref{psi}). The Lagrangian for $\gamma D_{(s)}^{(*)}D_{(s)}^{(*)}$ vertices is obtained with the help of gauge invariance \cite{Dong:2009uf},
\begin{eqnarray}\label{gamma}
{\cal L}_{D_{(s)}D_{(s)}\gamma}&=&
ieA_\mu D^-\!\stackrel{\leftrightarrow}{\partial^{\,\mu}}\!D^++ieA_\mu D_s^-\!\stackrel{\leftrightarrow}{\partial^{\,\mu}}\!D_s^+, \nonumber\\
{\cal L}_{D^*_{(s)}D_{(s)}\gamma}&=&
\frac{e}{4} \epsilon^{\mu\nu\alpha\beta}F_{\mu\nu}
\bigg(g_{D^{*-}D^+\gamma} D^{*-}_{\alpha\beta}D^+ +g_{D^{*0}D^0\gamma} \bar{D}^{*0}_{\alpha\beta}D^0\bigg)  \nonumber\\
&&+g_{D_s^{*-}D_s^+\gamma} D^{*-}_{s\alpha\beta}D_s^+ + \rm H.c. ,\nonumber \\
{\cal L}_{D_{(s)}^*D_{(s)}^*\gamma}&=&-ieA_\mu\bigg(g^{\alpha\beta}\!
D^{*-}_\alpha \!\stackrel{\leftrightarrow}{\partial^{\,\mu}}
 D^{*+}_\beta  \nonumber\\
&&-g^{\mu\beta} D^{*-}_\alpha \partial^\alpha D^{*+}_\beta
+ g^{\mu\alpha}\partial^\beta D^{*-}_\alpha D^{*+}_\beta\bigg) \nonumber\\
&&-ieA_\mu\bigg(g^{\alpha\beta}\!
D^{*-}_{s\alpha} \!\stackrel{\leftrightarrow}{\partial^{\,\mu}}
 D^{*+}_{s\beta}  \nonumber\\
&&-g^{\mu\beta} D^{*-}_{s\alpha} \partial^\alpha D^{*+}_{s\beta}
+ g^{\mu\alpha}\partial^\beta D^{*-}_{s\alpha} D^{*+}_{s\beta}\bigg),
\end{eqnarray}
where $A\stackrel{\leftrightarrow}{\partial}_{\mu} B = A \partial_{\mu} B - \partial_{\mu} A B$,
$F_{\mu\nu}=\partial_{\mu}A_{\nu}-\partial_{\nu}A_{\mu}$, and
$D^*_{\alpha\beta}=\partial_{\alpha}D^*_{\beta}-\partial_{\beta}D^*_{\alpha}$. Note that the electromagnetic interactions of $D^0D^0\gamma$ and $D^{*0}D^{*0}\gamma$ do not exist. Then the effective Lagrangians involved are
\begin{eqnarray}\label{psi}
\hspace*{-2cm}
{\cal L}_{\psi_n D^{(*)}D^{(*)}}&=&-ig_{\psi_n DD}
\psi_n^{\mu}D^{\dagger}
\!\stackrel{\leftrightarrow}{\partial}_{\mu} D
\nonumber \\
&&+g_{\psi_n DD^*} \epsilon_{\mu\nu\alpha\beta}
\partial^\mu\psi_n^{\nu} \bigg(
 D^{*\beta \dagger} \!\stackrel{\leftrightarrow}{\partial}^{\alpha}\!D-D^{\dagger}
\!\stackrel{\leftrightarrow}{\partial}^{\alpha}\!D^{*\beta}
\bigg)\nonumber\\
&&+ig_{\psi_n D^*D^*}
\psi_n^{\mu}\bigg(
-\bar{D}^{*\dagger\alpha}\!\stackrel{\leftrightarrow}{\partial}_{\mu}D^{*\dagger}_{\alpha}\nonumber\\
&&+\bar{D}_{\mu}^{*\dagger}\!\stackrel{\leftrightarrow}{\partial}_{\alpha}D^{*\dagger\alpha}
+\bar{D}^{*\dagger\alpha}\!\stackrel{\leftrightarrow}{\partial}_{\alpha}D^{*\dagger}_{\mu}
\bigg),\nonumber\\
{\cal L}_{\eta_c D^{(*)}_{(s)}D^{(*)}_{(s)}}&=&ig_{\eta_c D^*D}
D^{*\mu}\bigg(\partial_\mu\eta_cD^\dagger-\eta_c\partial_\mu D^\dagger\bigg)+ \rm H.c. \nonumber\\
&&-g_{\eta_c D^*D^*}\epsilon^{\mu\nu\alpha\beta}\partial_\mu D^*_{\nu}D^{*\dagger}_{\alpha}\partial_\beta\eta_c\nonumber\\
&&+ig_{\eta_c D_s^*D_s}
D_s^{*\mu}\bigg(\partial_\mu\eta_cD_s^\dagger-\eta_c\partial_\mu D_s^\dagger\bigg)+\rm H.c. \nonumber\\
&&-g_{\eta_c D_s^*D_s^*}\epsilon^{\mu\nu\alpha\beta}\partial_\mu D^*_{s\nu}D^{*\dagger}_{s\alpha}\partial_\beta\eta_c.
\end{eqnarray}
In the heavy quark limit, the coupling constant involved in the above Lagrangians $g_{\psi_n D^{(*)}D^{(*)}}$ and $g_{\eta_c D^{*}D^*}$ can be related to the gauge coupling $g_2$. One can determine $g_2=\sqrt{m_{\psi n}}/(2m_D f_\psi)$ by the vector meson dominance model \cite{Lin:1999ad,Colangelo:2003sa}, where $f_{\psi_n}$ can be obtained from the lepton decay widths of $J/\psi$ and $\psi(2S)$ \cite{Dong:2009uf,Chen:2015igx}. The coupling constants among the charmonia and charmed mesons satisfy the following relations \cite{Colangelo:2003sa},
\begin{eqnarray}
g_{\psi_n DD}&=& 2g_2\sqrt{m_{\psi_n}}m_D,             \nonumber\\
g_{\psi_n D^*D}&=& 2g_2\sqrt{ m_D m_{D^*}/m_{\psi_n}},   \nonumber\\
g_{\psi_n D^*D^*}&=& 2g_2\sqrt{m_{\psi_n}}m_D^*,      
\end{eqnarray}
and
\begin{eqnarray}\label{relation}
g_{\eta_c D^*D}&=& 2g_2\sqrt{m_{\eta_c}m_{D}m_{D^*}},           \nonumber\\
g_{\eta_c D^*D^*}&=& 2g_2m_{D^*}\sqrt{m_{\eta_c}} .   
\end{eqnarray}
The couplings $g_{\eta_c D_s D_s^*}$ and $g_{\eta_c D_s^*D_s^*}$ satisfy the same relation as in the Eq.~(\ref{relation}). The couplings $g_{D^{*0}D^0\gamma}$ and $g_{D^{*-}D^+\gamma}$ are fixed by data on the radiative decay widths of $D^{*0}$ and $D^{*+}$ and they are 2.0 $\rm{GeV^{-1}}$ and $-0.5$ $\rm{GeV^{-1}}$ \cite{Dong:2009uf}, respectively. The coupling $g_{D_s^{*}D_s\gamma}=(-0.3\pm0.1)$ $\rm{GeV^{-1}}$ is adopted from the QCD sum rules \cite{Becirevic:2009xp}. The relevant coupling constants are listed in Table \ref{coupling}.

\begin{table}[tbp]\centering
\caption{The relevant couplings in these calculations are listed.}
\label{coupling}
\renewcommand\arraystretch{1.05}
\begin{tabular*}{86mm}{@{\extracolsep{\fill}}m{1.47cm}m{0.9cm}m{1.47cm}m{1.47cm}m{1.47cm}}
\toprule[1pt]
\toprule[1pt]
Couplings                  &$g_{J/\psi DD}$           &$g_{J/\psi D^*D}$           &$g_{J/\psi D^*D^*}$        &$g_{\psi(2S) DD}$  \\
Values                    &7.44                      &2.49 $\rm{GeV^{-1}}$        &8.0                        &12.39              \\
$g_{\psi(2S) D^*D}$       &$g_{\psi(2S) D^*D^*}$     &$g_{\eta_c(1S ) DD^*}$       &$g_{\eta_c(1S) D^*D^*}$    &$g_{\eta_c(2S) DD^*}$\\
3.49 $\rm{GeV^{-1}}$      &13.33                     &7.58                        &2.63 $\rm{GeV^{-1}}$       &12.78\\
$g_{\eta_c(2S) D^*D^*}$   &$g_{\eta_c(1S) D_sD_s^*}$ &$g_{\eta_c(1S) D_s^*D_s^*}$ &$g_{\eta_c(2S) D_sD_s^*}$  &$g_{\eta_c(1S) D_s^*D_s^*}$   \\
3.64 $\rm{GeV^{-1}}$      &6.51                      &2.76 $\rm{GeV^{-1}}$        &14.44                       &3.83 $\rm{GeV^{-1}}$    \\
\bottomrule[1pt]
\bottomrule[1pt]
\end{tabular*}
\end{table}

The contact interaction is also required for the scattering process. The $D\bar{D}^{*}\gamma\psi_{n}$ four-point vertex can be obtained from the Lagrangian for $\psi_n DD^*$ interaction by substituting $\partial_\mu\rightarrow \partial_\mu+ iQeA_{\mu}$. Similarly, other contact Lagrangians can be obtained.

\subsection{Scattering amplitudes of $D\bar{D}^*$ with $J^{PC}=1^{++}$ to $\gamma\psi_n$}

Based on the Feynman rules and the given Lagrangians, the scattering amplitudes of $D(p_1)\bar{D}^*(p_2)\rightarrow\gamma(k_1) J/\psi(k_2)$ can be written as
\begin{eqnarray}
i\mathcal{M}^{(a)}_N&=&i^2\left[\frac{e}{4}g_{D^{*0}D^0\gamma}\varepsilon_{\mu\nu\alpha\beta} (ik_{1}^\mu g^\nu_{\lambda}-ik_{1}^\nu g^\mu_{\lambda}) (-ip_{2}^\alpha g^\beta_{\tau} \right. \nonumber\\
&&\left.+ip_{2}^\beta g^\alpha_{\tau})\epsilon_{\gamma}^{*\lambda}\epsilon_{\bar{D}^{*0}}^\tau\right]\left[-ig_{\psi_nDD}(-ip_{1\sigma}-iq_\sigma)
\epsilon^{*\sigma}_{\psi_n}\right]\nonumber\\
&&\times\frac{i}{q^2-m_{D^0}^2},
\end{eqnarray}

\begin{eqnarray}
i\mathcal{M}^{(b)}_N&=&i^2\left[\frac{e}{4}g_{D^{*0}D^0\gamma}\varepsilon_{\mu\nu\alpha\beta} (ik_{1}^\mu g^\nu_{\lambda}-ik_{1}^\nu g^\mu_{\lambda})(-iq^\alpha g^{\beta\rho}\right. \nonumber\\
&&\left. +iq^\beta g^{\alpha\rho})\epsilon_\gamma^{*\lambda}\right]ig_{\psi_nD^*D^*} \left[-(-iq-ip_{2})_\sigma g^\xi_{\tau} \right.\nonumber\\
&& \left.+(-iq-ip_2)_\tau g_{\sigma}^\xi+(-iq-ip_2)^\xi g_{\sigma\tau}\right] \epsilon_{\psi_n}^{*\sigma}\epsilon_{\bar{D}^{*0}}^\tau     \nonumber\\
&&\times\frac{i\left(-g_{\rho\xi}+q_\rho q_\xi/m^2_{\bar{D}^{*0}}\right)}{q^2-m^2_{\bar{D}^{*0}}},
\end{eqnarray}

\begin{eqnarray}
i\mathcal{M}^{(c)}_C&=&i^2\left[ie(-ip_{1\lambda}+iq_\lambda)\epsilon^{*\lambda}_\gamma\right] \left[g_{\psi_nDD^*}\varepsilon_{\mu\sigma\alpha\tau}(ik_{2}^\mu)      \right.  \nonumber\\ &&\times\left. (iq^\alpha+ ip_2^\alpha)\epsilon_{\psi_n}^{*\sigma}\epsilon_{D^{*-}}^\tau \right]\frac{i}{q^2-m_{D^-}^2},
\end{eqnarray}
\begin{eqnarray}
i\mathcal{M}^{(d)}_C&=&i^2\left[\frac{e}{4}g_{D^{*-}D^+\gamma}\varepsilon_{\mu\nu\alpha\beta} (ik_{1}^\mu g^\nu_{\lambda}-ik_{1}^\nu g^\mu_{\lambda}) (-ip_{2}^\alpha g^\beta_{\tau}  \right. \nonumber\\
&&\left.+ip_{2}^\beta g^\alpha_{\tau})\epsilon_{\gamma}^{*\lambda}\epsilon_{\bar{D}^{*-}}^\tau\right]\left[-ig_{\psi_nDD}(-ip_{1\sigma}-iq_\sigma)
\epsilon^{*\sigma}_{\psi_n}\right]\nonumber\\
&&\times\frac{i}{q^2-m_{D^+}^2},
\end{eqnarray}
\begin{eqnarray}
i\mathcal{M}^{(e)}_C&=&i^2\left[\frac{e}{4}g_{D^{*-}D^+\gamma}\varepsilon_{\mu\nu\alpha\beta} (ik_{1}^\mu g^\nu_{\lambda}-ik_{1}^\nu g^\mu_{\lambda})(-iq^\alpha g^{\beta\rho}\right. \nonumber\\
&&\left. +iq^\beta g^{\alpha\rho})\epsilon_\gamma^{*\lambda}\right]ig_{\psi_nD^*D^*} \left[-(-iq-ip_{2})_\sigma g^\xi_{\tau} \right.\nonumber\\
&& \left.+(-iq-ip_2)_\tau g_{\sigma}^\xi+(-iq-ip_2)^\xi g_{\sigma\tau}\right] \epsilon_{\psi_n}^{*\sigma}\epsilon_{{D}^{*-}}^\tau     \nonumber\\
&&\times\frac{i\left(-g_{\rho\xi}+q_\rho q_\xi/m^2_{{D}^{*-}}\right)}{q^2-m^2_{{D}^{*-}}},
\end{eqnarray}
\begin{eqnarray}
i\mathcal{M}^{(f)}_C&=&i^2(-ie)\left[(-iq_\lambda+ip_{2\lambda})g_{\tau\rho}-(-iq_\tau g_{\lambda\rho})-ip_{2\rho}g_{\lambda\tau} \right]  \nonumber\\
&&\times \epsilon_{D^{*-}}^\tau\epsilon_\gamma^{*\lambda} \left[ g_{\psi_nDD^*}\varepsilon_{\mu\sigma\alpha\xi}(ik_{2\mu})(-ip_{1\alpha}-iq_\alpha) \epsilon_{\psi_n}^{*\sigma}\right] \nonumber\\
&&\times\frac{i\left(-g^{\rho\xi}+q^\rho q^\xi/m_{D^{*+}}^2\right)}{q^2-m_{D^{*+}}^2},
\end{eqnarray}
\begin{eqnarray}\label{con}
i\mathcal{M}^{(g)}_{C}&=&i2eg_{\psi_nDD^*}\varepsilon_{\mu\nu\alpha\beta}(k_{2}^\mu)\epsilon_{\psi_n}^{*\nu}\epsilon_\gamma^{*\alpha}\epsilon_{D^*}^\beta,
\end{eqnarray}
where $q$ is the four-momentum of the exchanged charmed meson, and the polarization vectors $\epsilon^\tau_{D^*}$, $\epsilon^{*\lambda}_{\gamma}$ and $\epsilon^{*\sigma}_{\psi_n}$ are for the $D^{*}$ charmed meson, photon and $\psi_{n}$ meson, respectively. The superscripts (a$-$g) correspond to the labels of the individual diagrams for $D\bar{D}^{*}\rightarrow\gamma \psi_n$ in Fig. \ref{tree}. The details of the polarization vectors are given in the Appendix. The subscripts $C$ and $N$ denote the corresponding amplitudes from the charged and neutral channels, respectively. Thus, the scattering amplitudes of $D\bar{D}^{*}\rightarrow\gamma \psi_n$ are
\begin{eqnarray}\label{total width}
\mathcal{M}_{N}&=&\mathcal{M}^{(a)}_N+\mathcal{M}^{(b)}_N,\nonumber\\
\mathcal{M}_{C}&=&\mathcal{M}^{(c)}_C+\mathcal{M}^{(d)}_C+\mathcal{M}^{(e)}_C+\mathcal{M}^{(f)}_C+\mathcal{M}^{(g)}_{C}.   
\end{eqnarray}
We introduce $\mathcal{M}_{N}^{[m_s]}$ and $\mathcal{M}_{C}^{[m_s]}$ especially for the $z$-component of the $D^*$ spin, which is $m_s$.

With the above amplitudes, using Case \Rmnum{3} of the $X(3872)$ as an example, the decay amplitude can be written as
\begin{eqnarray}\label{decay ampltitude}
&&\mathcal{M}^{1, M}_{{X(3872)}\rightarrow \gamma\psi_n}
=\sqrt{2}\frac{\sqrt{2m_{X(3872)}}}{\sqrt{2m_D}\sqrt{2m_{D^*}}}\int\frac{\rm  d^3\textbf{p}}{(2\pi)^{3/2}}\,\mathcal{F}^2(\overline{\textbf{p}}^2) \nonumber\\
&&\qquad\quad \times  \left(\sum_{m_s}\phi_{[D^0\bar{D}^{*0}]|^3S_1\rangle}C^{1,M}_{1m_s,00}\,\mathcal{M}_{N}^{[m_s]} 
\right.\nonumber\\&&\qquad\quad\left.
+\sum_{m_s,m_L}\phi_{[D^0\bar{D}^{*0}]|^3D_1\rangle}C^{1,M}_{1m_s,2m_L}Y_{2,m_L}\mathcal{M}_{N}^{[m_s]} \right.\nonumber\\
&&\qquad\quad\left.+\sum_{m_s}\phi_{[D^+D^{*-}]|^3S_1\rangle}C^{1,M}_{1m_s,00}\mathcal{M}_{C}^{[m_s]}   \right.\nonumber\\
&&\qquad\quad\left.+\sum_{m_s,m_L}\phi_{[D^+D^{*-}]|^3D_1\rangle}C^{1,M}_{1m_s,2m_L}Y_{2,m_L}\mathcal{M}_{C}^{[m_s]} \right) .  
\end{eqnarray}
In the above amplitudes, $\mathcal{F}(\overline{\textbf{p}}^2)=\exp(-\overline{\textbf{p}}^{2n}/\Lambda_\alpha^{2n})$ is the form factor with Gaussian form to suppress the large momentum contributions and $\overline{\textbf{p}}$ is the 3-momentum difference between the initial and final particles. 

We do not explicitly list the contribution of $D^* \bar D$ in Eq. (\ref{decay ampltitude}) since it can be related to that from $D \bar D^*$ by charge conjugation. The first factor $\sqrt{2}$ in Eq. (\ref{decay ampltitude}) is due to these conjugated diagrams. Now the radiative decay width can be obtained by combining Eq. (\ref{decay ampltitude}) and Eq. (\ref{width1}). 

\subsection{Scattering amplitudes of the $X(3872)$ partners}

The $D\bar{D}^*$ with $J^{PC}=1^{+-}$ can decay to $\gamma\eta_c(1S)$ and $\gamma\eta_c(2S)$. Compared to the $X(3872)$, the $D^*\bar{D}^*$ would couple to the negative $C$-parity state via the $S$-wave interaction and thus more diagrams appear in Fig. \ref{partnerfig}. 

The scattering amplitudes for process ${D}^{(*)}(p_1)\bar D^*(p_2)\rightarrow\gamma(k_1)\eta_c(k_2)$ are
\begin{eqnarray}\label{1+-a}
i\mathcal{M}^{a}_N&=&i^2\left[\frac{e}{4}g_{D^{*0}D^0\gamma}\varepsilon_{\mu\nu\alpha\beta}(ik_1^\mu g_\lambda^\nu-ik_1^\nu g_\lambda^\mu)(-iq^{\alpha}g^{\beta\rho}          \right. \nonumber\\
&&\left.+iq^{\beta}g^{\alpha\rho})\epsilon_{\gamma}^{*\lambda}\right] \left[-g_{\eta_{c}D^*D^*}\varepsilon_{\sigma\xi\tau\delta}(iq^{\sigma})(ik_{2}^{\delta})\epsilon_{\bar{D}^{*0}}^{\tau} \right]\nonumber\\
&&\times\frac{i\left(-g_{\rho}^{\xi}+q_{\rho}q^{\xi}/m^2_{\bar{D}^{0*}}\right)}{q^2-m_{\bar{D}^{0*}}^2},
\end{eqnarray}
\begin{eqnarray}
i\mathcal{M}^{b}_C&=&i^2\left[ie(-ip_{1\lambda}+iq_\lambda)\epsilon^{*\lambda}_\gamma\right]
\left[ig_{\eta_cDD^*}(ik_{2\tau}-iq_\tau)\epsilon_{D^{*-}}^{\tau}\right]    \nonumber\\
&&\times\frac{i}{q^2-m_{D^-}^2},
\end{eqnarray}
\begin{eqnarray}
i\mathcal{M}^{c}_C&=&i^2\left[\frac{e}{4}g_{D^{*-}D^+\gamma}\varepsilon_{\mu\nu\alpha\beta}
(ik_1^\mu g_\lambda^\nu-ik_1^\nu g_\lambda^\mu)(-iq^{\alpha}g^{\beta\rho}   \right. \nonumber\\
&&\left.+iq^{\beta}g^{\alpha\rho})\epsilon_{\gamma}^{*\lambda}\right]\left[-g_{\eta_{c}D^*D^*}\varepsilon_{\sigma\xi\tau\delta}
(iq^{\sigma})(ik_{2}^{\delta})\epsilon_{D^{*-}}^{\tau} \right] \nonumber\\
&&\times\frac{i\left(-g_{\rho}^{\xi}+q_{\rho}q^{\xi}/m_{D^{*-}}^2\right)}{q^2-m_{D^{*-}}^2},
\end{eqnarray}
\begin{eqnarray}
i\mathcal{M}^{d}_C&=&i^2(-ie)\left[(-iq_\lambda+ip_{2\lambda})g^{\rho}_{\tau} -(-iq_{\tau})g^{\rho}_{\lambda} -ip_{2}^\rho g_{\lambda\tau}\right]  \nonumber \\
&&\times\epsilon_{D^{*-}}^\tau\epsilon_\gamma^{*\lambda}\left[ig_{\eta_cDD^*}(ik_{2}^\xi+ip_{1}^\xi)\right]\nonumber\\
&&\times\frac{i\left(-g_{\rho\xi}+q_\rho q_\xi/m_{D^{*+}}^2\right)}{q^2-m_{D^{*+}}^2},
\end{eqnarray}
\begin{eqnarray}
i\mathcal{M}^{e}_{C}&=&-ieg_{\eta_cDD^*}\epsilon_\gamma^{*\mu}\epsilon_{\mu D^{*-}},
\end{eqnarray}
\begin{eqnarray}
i\mathcal{M}^{f}_N&=&i^2\left[\frac{e}{4}g_{D^{*0}D^0\gamma}\varepsilon_{\mu\nu\alpha\beta}(ik_{1}^{\mu} g^\nu_{\lambda}-ik_{1}^\nu g_{\lambda}^\mu)(-ip_1^\alpha g^{\beta}_\tau+ip_1^\beta \right.\nonumber\\
&&\left.\times g^\alpha_{\tau})\epsilon^{*\lambda}_{\gamma}\epsilon^{\tau}_{\bar{D}^{*0}}\right]\left[ig_{\eta_cD^*D} (ik_{2\sigma}-iq_\sigma)\epsilon_{D^{*0}}^{\sigma}\right]\nonumber\\
&&\times\frac{i}{q^2-m^2_{D^0}},
\end{eqnarray}
\begin{eqnarray}
i\mathcal{M}^{g}_N&=&i^2\left[\frac{e}{4}g_{D^{*0}D^0\gamma}\varepsilon_{\mu\nu\alpha\beta}(ik_{1}^\mu g_\lambda^\nu-ik_1^\nu g_\lambda^\mu)(-ip_{2}^\alpha g^\beta_{\sigma} +ip^\beta_{2} \right. \nonumber\\
&&\times\left.g^\alpha_{\sigma}) \epsilon^{*\lambda}_{\gamma}\epsilon^{\sigma}_{D^{*0}}\right]  \left[ig_{\eta_cD^*D}(ik_{2\tau}-iq_\tau)\epsilon_{\bar{D}^{*0}}^{\tau}\right] \nonumber\\ &&\times\frac{i}{q^2-m^2_{D^0}},
\end{eqnarray}
\begin{eqnarray}
i\mathcal{M}^{h}_C&=&i^2\left[\frac{e}{4}g_{D^{*-}D^+\gamma}\varepsilon_{\mu\nu\alpha\beta}(ik_{1}^{\mu} g^\nu_{\lambda}-ik_{1}^\nu g_{\lambda}^\mu)(-ip_1^\alpha g^{\beta}_\tau    \right.\nonumber\\
&&\left.-(-ip_1^\beta) g^\alpha_{\tau})\epsilon^{*\lambda}_{\gamma}\epsilon^{\tau}_{D^{*-}}\right]\left[ig_{\eta_cD^*D} (ik_{2\sigma}-iq_\sigma))\epsilon_{D^{*+}}^{\sigma}\right]\nonumber\\
&&\times\frac{i}{q^2-m^2_{D^+}},
\end{eqnarray}
\begin{eqnarray}
i\mathcal{M}^{i}_C&=&i^2\left[\frac{e}{4}g_{D^{*+}D^-\gamma}\varepsilon_{\mu\nu\alpha\beta}(ik_{1}^\mu g_\lambda^\nu-ik_1^\nu g_\lambda^\mu)(-ip_{2}^\alpha g^\beta_{\sigma}  \right. \nonumber\\
&&+ip^\beta_{2}\left.g^\alpha_{\sigma}) \epsilon^{*\lambda}_{\gamma}\epsilon^{\sigma}_{D^{*+}}\right]  \left[ig_{\eta_cD^*D}(ik_{2\tau}-iq_\tau)\epsilon_{D^{*-}}^{\tau}\right] \nonumber\\ &&\times\frac{i}{q^2-m^2_{D^-}},
\end{eqnarray}
\begin{eqnarray}
i\mathcal{M}^{j}_C&=&i^2(-ie)\left[(-iq_\lambda +ip_{1\lambda})g_{\tau\rho}+iq_\tau g_{\rho\lambda}-ip_{1\rho}g_{\tau\lambda} \right] \nonumber\\
&&\times\epsilon_{D^{*-}}^\tau\epsilon_\gamma^{*\lambda}\left[-g_{\eta_cD^*D^*}
\varepsilon_{\mu\sigma\xi\beta}(-ip_2^\mu)(ik_{2}^\beta)\epsilon_{D^{*+}}^\sigma\right]  \nonumber\\
&&\times\frac{i\left(-g^{\rho\xi}+q^\rho q^\xi/m_{D^{*+}}^2\right)}{q^2-m_{D^{*+}}^2},
\end{eqnarray}
\begin{eqnarray}
i\mathcal{M}^{k}_C&=&i^2(-ie)\left[(-ip_{2\lambda}+iq_\lambda)g_{\sigma\rho}
+ip_{2\rho}g_{\lambda\sigma}-iq_\sigma g_{\rho\lambda} \right] \nonumber\\
&&\times\epsilon_{D^{*+}}^\sigma\epsilon_\gamma^{*\lambda}
\left[-g_{\eta_cD^*D^*}\varepsilon_{\mu\xi\tau\beta}(iq_{\mu})(ik_{2\beta}) \epsilon_{D^{*-}}^\tau\right] \nonumber\\
&&\times\frac{i\left(-g^{\rho\xi}+q^\rho q^\xi/m_{D^{*+}}^2\right)}{q^2-m_{D^{*-}}^2},
\end{eqnarray}
\begin{eqnarray}\label{1+-l}
i\mathcal{M}^{l}_{C}&=&ieg_{\eta_cD^*D^*}\varepsilon_{\mu\tau\sigma\beta}k_{2}^\beta\epsilon_\gamma^{*\mu}\epsilon_{D^{*-}}^\tau
\epsilon_{D^{*+}}^\sigma.
\end{eqnarray}
Here, the superscripts in the amplitudes $\mathcal{M}^{a-l}$ for $D^{(*)}\bar{D}^{*}~(1^{+-})\rightarrow\gamma \eta_c$ correspond to the labels of the individual diagrams in Fig. \ref{partnerfig}. 

According to the flavor symmetry, the scattering amplitudes of $D_s\bar{D}_s^*$ with $J^{PC}=1^{+-}$ decaying to $\gamma\eta_c$ can be obtained by the following substitutions,
\begin{eqnarray}
\mathcal{M}_{\rm tot}&=&( \mathcal{M}^{b}_C+\mathcal{M}^{c}_C+\mathcal{M}^{d}_C+\mathcal{M}^{e}_{C}
)\left|^{ m_{D^{(*)}} \rightarrow m_{D_s^{(*)} },\,
g_{D^{*-}D^+\gamma}\rightarrow g_{D_s^*D_s\gamma}
}_{ 
\epsilon_{D^*}\rightarrow \epsilon_{D^*_s},\,
g_{\eta_{c}D^{(*)}D^*}\rightarrow g_{\eta_{c}D_s^{(*)}D_s^*}
}\right. .\nonumber\\
\end{eqnarray}

With the above expressions, we can obtain their radiative decay widths, the same as for the $X(3872)$.

\section{Numerical results for radiative decay}\label{sec5}
In this section, we present numerical results for  radiative decays of the $X(3872)$ and its partner states.  

Gauge invariance requires that for any QED process involving an external photon with momentum $k_1$, the amplitude $\epsilon_{\gamma}^{*\mu}\mathcal{M}_\mu$ vanishes when we replace $\epsilon_{\gamma}^{*\mu}$ with $k_1^{\mu}$, i.e., $k_1^{\mu}\mathcal{M}_\mu=0$. We have checked that our framework respects this symmetry. Although we only show the tree diagrams, in Eq. (\ref{amplitude}) there is a momentum integral which also appears in the usual triangle-loop calculation in some models where the $X(3872)$ is treated as a point particle and strongly couples to $D\bar D^{(*)}$. Gauge invariance is easy to maintain in our approach since the momentum integral is outside of the scattering amplitudes $D\bar D^{(*)}\to \gamma \psi_n$.     

Rotation symmetry suggests that the decay width of an unstable particle is independent of the polarization of the initial particle. Thus, for the spin-1 particles, the radiative decay width should satisfy the relation $\Gamma_{[AB]\rightarrow CD}^{1,1} =\Gamma_{[AB]\rightarrow CD}^{1,0}=\Gamma_{[AB]\rightarrow CD}^{1,-1}$. Our results have this relation. Alternatively, this symmetry can be used to simplify the calculation in Eq. (\ref{width1}) as 
\begin{equation}\label{width2}
\begin{split}
\Gamma_{[AB] \rightarrow CD}^{J,M}=\frac{|\textbf{k}|}{32\pi^2m_{[AB]}^2}\sum_{M'}\int& {\rm d}\Omega_{\textbf{k}}D^{J}_{M M'}D^{*J}_{M M'}(-\phi_{\textbf{k}}, -\theta_{\textbf{k}}, 0)\\
& \times\left|\mathcal{M}^{JM'}_{[AB]\rightarrow CD}\right|_{\phi_{\textbf{k}}=\theta_{\textbf{k}}=0}^2,
\end{split}
\end{equation}
where we only need the scattering amplitudes with $\textbf{k}$ along the $z$ axis. $\mathcal{M}^{JM}_{[AB]\rightarrow CD}|_{\phi_{\textbf{k}}=\theta_{\textbf{k}}=0}$ is not zero only when $M=S_z(C)+S_z(D)$.

\subsection{Radiative decay width of the $X(3872)$}
The $X(3872)$ is a loosely bound state, so we calculate its radiative decay width in the range where its binding energy is less than 1 MeV.
Using the approach described in Sec. \ref{sec4}, we calculate the radiative decay width of two decay modes which are the $X(3872) \rightarrow \gamma J/\psi$ and the $X(3872) \rightarrow \gamma\psi(2S)$, and give the values of the ratio $R_{\gamma\psi}=\frac{\Gamma_{(X(3872) \rightarrow \gamma\psi(2S))}}{\Gamma_{(X(3872) \rightarrow \gamma J/\psi)}}$ within the pure molecular picture.

\begin{figure*}[tbp]
\includegraphics[width=505pt]{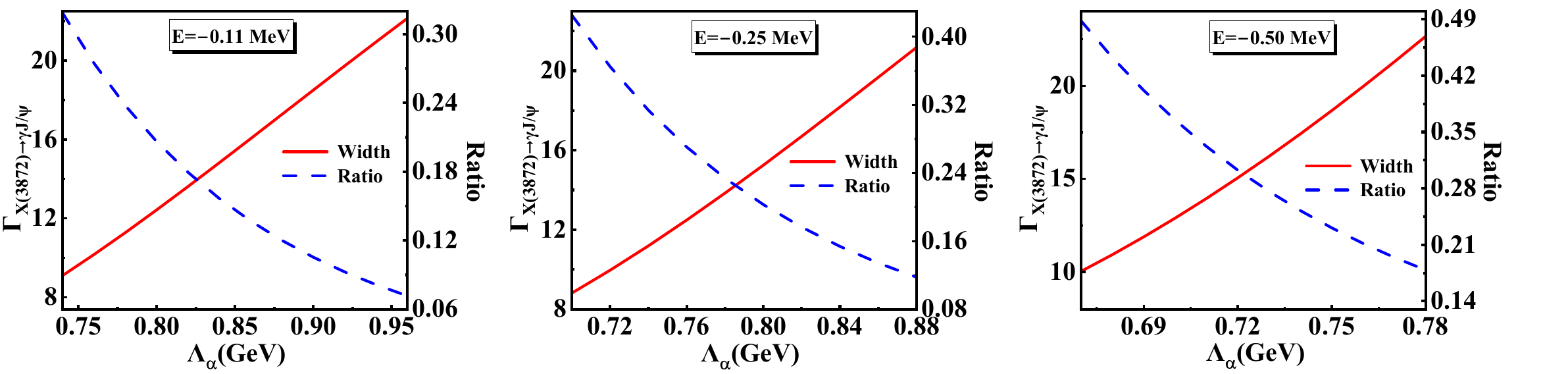}
\caption{The radiative decay widths of the $X(3872) \rightarrow \gamma J/\psi$ and the ratios $R_{\gamma\psi}=\frac{\Gamma_{X(3872) \rightarrow \gamma \psi(2S)}}{\Gamma_{X(3872) \rightarrow \gamma J/\psi}}$ for three typical binding energies $E =-0.11$ MeV, $-0.25$ MeV, $-0.50$ MeV. The units of the radiative decay width are keV.}\label{X(3872)width51}
\end{figure*}

According to the results measured by the BaBar \cite{BaBar:2008flx} and Belle \cite{Belle:2011wdj} Collaborations and combining the width $\Gamma_X=1.19\pm0.21$ MeV in the PDG \cite{ParticleDataGroup:2022pth}, one can get the radiative decay width of the $X(3872) \rightarrow \gamma J/\psi$,
 \begin{equation}
 {\Gamma_{X(3872) \rightarrow \gamma J/\psi}}=\left\{
     \begin{array}{lr}
        10.1^{+4.6}_{-4.5}~\rm keV           &\rm Belle\\
       15.5\pm7.3~\rm keV   &\rm BaBar
     \end{array}
\right..
\end{equation}
We present our numerical results for three typical binding energies $E =-0.11$ MeV, $-0.25$ MeV, $-0.50$ MeV in Fig. \ref{X(3872)width51}.
For the case of the binding energy $E=-0.11$ MeV, when the cutoff $\Lambda_\alpha$ is around 0.75 GeV$-$0.95 GeV, the radiative decay width of the $X(3872)$ decay to $\gamma J/\psi$ is in the range of 10 keV$-$20 keV. It is clear that the theoretical width and the experimental measurements are consistent within the error. 

In the context of the molecular state, the ratio $R_{\gamma\psi}$ is small as seen in Fig. \ref{X(3872)width51} and the process of $X(3872) \rightarrow \gamma\psi(2S)$ is thus suppressed. The ratio increases as the binding energy $|E|$ increases. However, the upper limit of the ratio $R_{\gamma\psi}$ is about 0.5 with the reasonable cutoff $\Lambda_\alpha$ set by the $\Gamma_{X(3872) \rightarrow \gamma J/\psi}$ in this work.  The ratio $R_{\gamma\psi}$ measured by different collaborations is different
\begin{equation}
 R_{\gamma\psi}\,\left\{
     \begin{array}{lr}
        =3.4\pm1.4~(3.5\sigma)           &\rm{BaBar}\\
        =2.46\pm0.64\pm0.29~(4.4\sigma)   &\rm{LHCb}\\
        \textless~2.1~(90\%~\rm{C.L.})                    &\rm{Belle}\\
       \textless~0.59~(90\%~\rm{C.L.})                    & \rm{BES\Rmnum{3}}\nonumber
     \end{array}
\right..
\end{equation} 
Note that the BaBar and LHCb measurements challenge the results of BES\Rmnum{3} and Belle, but the experimental uncertainties are also not small either\cite{BaBar:2008flx,LHCb:2014jvf,Belle:2011wdj,BESIII:2020nbj}. A predominantly molecular nature of the $X(3872)$ in our calculations is more compatible with the measurements of BES\Rmnum{3} and Belle.

Such a large ratio $R_{\gamma\psi}$ from BaBar and LHCb cannot be naturally explained either in Ref. \cite{Swanson:2004pp}, where the $X(3872)$ is assumed to be a $D\bar{D}^*$ molecule mixed with small $\rho J/\psi$ and $\omega J/\psi$. The $D\bar{D}^*$ components decay to a photon via light-quark annihilation, while the $\rho J/\psi$ and $\omega J/\psi$ decay to a photon via the vector meson dominance. The predicted ratio $R_{\gamma\psi}$ is about 4$\times10^{-3}$.

The model of the molecule-charmonium mixture gave that $R_{\gamma\psi}$ is about 0.5$-$5 \cite{Eichten:2005ga}. Ref. \cite{Dong:2009uf} suggested that the $X(3872)$ as a superposition of the molecular and charmonium components is necessary to explain the ratio $R_{\gamma\psi}$ by restricting to the central value of the ratio width $R_{\gamma\psi}=3.5$ in the potential model \cite{Swanson:2003tb}. The coupling channels including $D^0\bar{D}^{*0}$, $D^+D^{*-}$, $D_s^+D_s^{*-}$, $D^*D^*$ and $c\bar{c}$ are discussed in Ref. \cite{Cardoso:2014xda} where the branching ratio is 1.17. The molecule-charmonium mixture in the $X(3872)$ was used to explain the ratio $R_{\gamma\psi}$ in Refs. \cite{Cincioglu:2016fkm,Cincioglu:2019gzd}. In addition, the configuration of the pure charmonium state is also proposed to calculate $R_{\gamma\psi}$ and the ratio is about 1.2$-$15 \cite{Badalian:2012jz,Barnes:2005pb,Barnes:2003vb,Li:2009zu,Lahde:2002wj,Mehen:2011ds,Wang:2010ej}.

According to our calculations, the radiative decay width is dominated by the $S$ wave and the contribution of the $D$ wave is very small. For the binding energy $E=-0.11$ MeV and the cutoff $\Lambda_\alpha=0.8$ GeV, the contribution from both $S$ and $D$ waves gives $\Gamma_{X(3872) \rightarrow \gamma J/\psi}=12.4$ keV and removing the contribution from the $D$ wave gives $\Gamma_{X(3872) \rightarrow \gamma J/\psi}=12.0$ keV. There is a difference of  0.4 keV between the two cases. Therefore, in the following calculations we do not consider the contribution of the $D$ wave anymore for the partner states. 

Since the hadrons are not pointlike particles, we usually introduce the form factor at each interaction vertices to describe the off-shell effect of the exchanged light mesons and to reflect the inner structure effect of the discussed hadrons. In general, $\mathcal{F}(q^2,m_E^2) = (\Lambda^2-m_E^2)/(\Lambda^2-q^2)$ is often used to discuss the interaction between two hadrons within the one-boson-exchange model. Such a formalism is a direct extension of the meson exchange model involved in the nuclear force. However, $\mathcal{F}(q^2,m_E^2)$ may reduce the strength of the vector meson exchange interactions compared to the pion exchange interaction. In the local hidden gauge approach, the interaction between two hadrons primarily arises from the exchange of the vector mesons. In fact, the difference between the one-boson-exchange model considering $\mathcal{F}(q^2,m_E^2)$ and the local hidden gauge approach reflects the different treatments of the relative weights of the meson exchanges. Despite all this, the one-boson-exchange model considering $\mathcal{F}(q^2,m_E^2)$ and the local hidden gauge approach can usually can give the similar conclusion for the binding energy of the hadronic molecular state \cite{Li:2012cs,Gamermann:2007fi,Chen:2019asm,Xiao:2019aya,Liu:2020nil,Molina:2020hde}

In the following discussions, we also take the form factor $\mathcal{F}(q^2) = \Lambda^2/(\Lambda^2-q^2)$ to discuss the bound state properties and the radiative decay widths of the $X(3872)$, and  $\mathcal{F}(q^2)$ can change the relative weight of the vector meson exchange with respect to the pion exchange compared to $\mathcal{F}(q^2,m_E^2)$. The Coulomb interaction can affect on the binding energy with the form factors being whether $\mathcal{F}(q^2,m_E^2)$ or $\mathcal{F}(q^2)$. With the binding energy $E=-0.11$ MeV in Case II, the binding energy $E$ becomes $-0.32$ MeV for $\mathcal{F}(q^2,m_E^2)$ while $-0.23$ MeV for $\mathcal{F}(q^2)$ after adding the Coulomb interaction. For the binding energy $E=-0.11$ MeV, the decay width of $X(3872)\to \gamma J/\psi$ and $R_{\gamma\psi}$ are in the ranges of around 8$-$20 keV and 0.1$-$0.3 for $\mathcal{F}(q^2,m_E^2)$ while 5$-$10 keV and 0.2$-$0.5 for $\mathcal{F}(q^2)$ with same choice of $\Lambda_\alpha$, respectively. One can notice that different form factors with the matched cutoffs lead to some differences in the Coulomb effects and the rate of electromagnetic production, due to different relative weights of the meson exchanges. Future experimental studies will be very useful to investigate such effects.

\subsection{Radiative decay widths of the $X(3872)$ partners}
\begin{figure}[tbp]
\includegraphics[width=250pt]{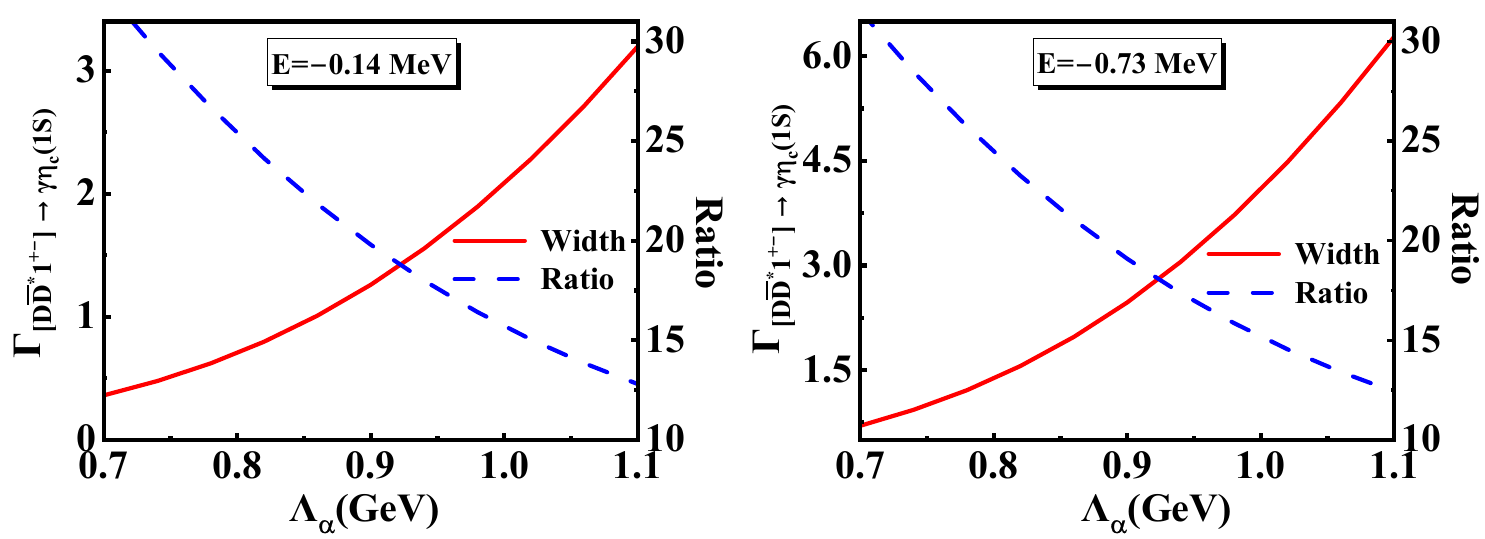}
\caption{The decay widths of $D\bar{D}^*$ with $J^{PC}=1^{+-}$ decaying to $\gamma \eta_c(1S)$ for two typical binding energies $E =-0.14$ MeV, $-0.73$ MeV and the ratios $R_{\gamma\eta_c}=\frac{\Gamma_{[D\bar{D}^*] \rightarrow \gamma \eta_c(2S)}}{\Gamma_{[D\bar{D}^*] \rightarrow \gamma \eta_c(1S)}}$. The units of the radiative decay width are keV.}\label{Cpartnerwidth}
\end{figure}

\begin{figure}[tbp]
\includegraphics[width=250pt]{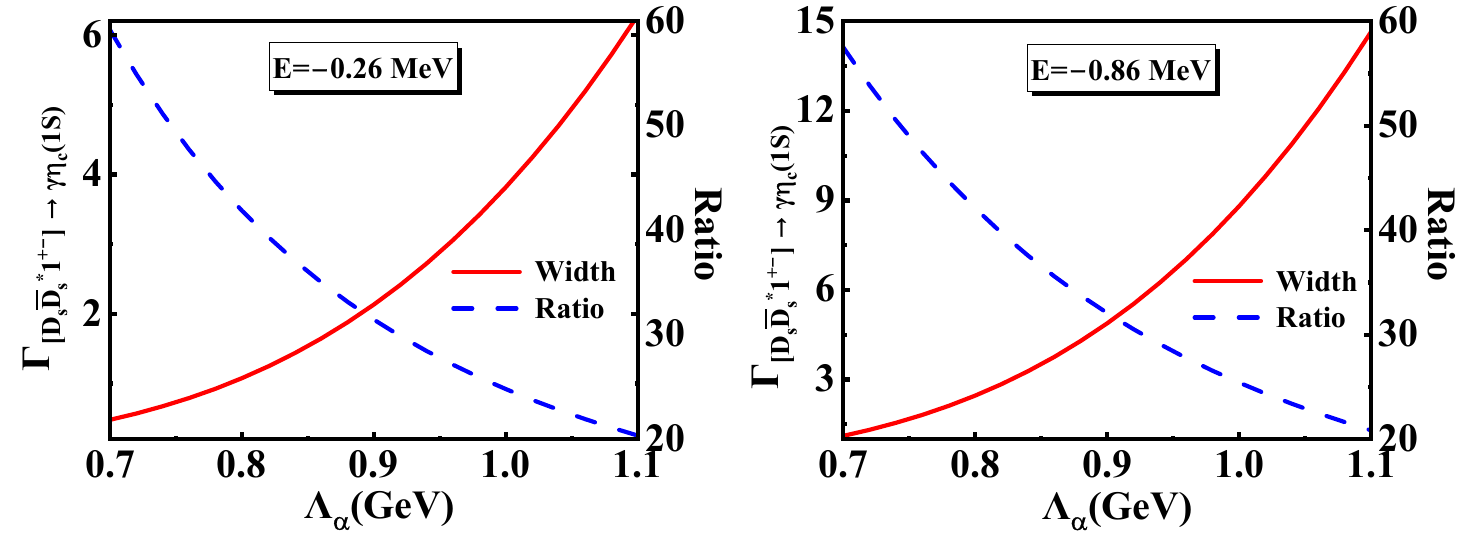}
\caption{The decay widths of $[D_s\bar{D}_s^*]$ with $J^{PC}=1^{+-}$ decaying to $\gamma \eta_c(1S)$ for two typical binding energies $E =-0.26$ MeV, $-0.86$ MeV and the ratios $R_{\gamma\eta_c}^s=\frac{\Gamma_{[D_s\bar{D}_s^*]\rightarrow \gamma \eta_c(2S)}}{\Gamma_{[D_s\bar{D}_s^*] \rightarrow \gamma \eta_c(1S)}}$. The units of radiative decay width are keV.}\label{Fpartnerwidth}
\end{figure}

We have obtained bound solutions of three partner states of the $X(3872)$ by solving the coupled channel Schr\"{o}dinger equation in Sec. \ref{sec3}. However, the $[D_s\bar{D}^*_s]$ with $J^{PC}=1^{++}$ is weakly bound compared to that with $J^{PC}=1^{+-}$, so we do not consider the radiative decay of the former in our calculations. The $[D_s\bar{D}^*_s]$ bound state with $J^{PC}=1^{+-}$ should exist with the SU(3) flavor symmetry and heavy quark spin symmetry in Ref. \cite{Meng:2020cbk}. Since the $[D\bar{D}^*]$ and $[D_s\bar{D}^*_s]$ states have the negative $C$ parity, it is very likely that they decay into a photon plus a pseudoscalar meson such as $\eta_c(1S)$ or $\eta_c(2S)$. Using the same method, we calculate two radiative decay modes that $[D_{(s)}\bar{D}^*_{(s)}]$ with $J^{PC}=1^{+-}$ decay to $\gamma\eta_c(1S)$ and $\gamma\eta_c(2S)$. The radiative decay width and ratio $R_{\gamma\eta_c}^{(s)}=\frac{\Gamma_{[D_{(s)}\bar{D}^*_{(s)}] \rightarrow \gamma \eta_c(2S)}}{\Gamma_{[D_{(s)}\bar{D}^*_{(s)}] \rightarrow \gamma \eta_c(1S)}}$ are shown in Fig. \ref{Cpartnerwidth} and Fig. \ref{Fpartnerwidth}. 

For the partner state $[D\bar{D}^*]$ with $J^{PC}=1^{+-}$, we present our numerical results with two typical binding energies $E =-0.14$ and $-0.73$ MeV in Fig. \ref{Cpartnerwidth}. It can be seen that the decay width in $\gamma\eta_c(2S)$ channel is much larger than that in the $\gamma\eta_c(1S)$ channel, which is obviously different from the case of the $X(3872)$. The ratio $R_{\gamma\eta_c}$ is about 10$-$30 as $\Lambda_\alpha$ changes from 0.7 GeV to 1.1 GeV. Ref. \cite{Ortega:2021yis} explored the possible existence of the $J^{PC}=1^{+-}$ counterpart of the $X(3872)$ as a mixing state of the charmonium and the molecule within a constituent quark model, and predicted the radiative width $\Gamma_{\gamma\eta_c(1S)}=69$ keV and $R_{\gamma\eta_c}=0.9$.

For the $D_s\bar{D}_s^*$ molecule with $J^{PC}=1^{+-}$, we show our numerical results for two binding energies $E = -0.26$ MeV and $-0.86$ MeV in Fig. \ref{Fpartnerwidth}. The width of the decay channel $\gamma\eta_c(1S)$ is very small, only a few keV, with the cutoff $\Lambda_\alpha$ in the range of 0.7$\sim$1.1 GeV. However, $R_{\gamma\eta_c}^s$ is large, suggesting that the width of decay to $\gamma\eta_c(2S)$ is easier to measure experimentally. The cutoff should be of the same order with heavy quark symmetry \cite{Lu:2014ina,Altenbuchinger:2013vwa}, and thus, the need of cutoffs larger than 2 GeV to bind means that the $D_s\bar{D}_s^*$ molecular states with $J^{PC}=1^{+\pm}$ most probably do not exist. There is still no experimental data on these states and radiative decay widths, and we look forward to the future experiments that can test our results.

\section{summary}\label{sec6}
Many exotic hadrons have been observed at modern facilities since 2003 \cite{Chen:2016spr,Brambilla:2019esw,Liu:2019zoy,Chen:2022asf}. As the first charmoniumlike state, the nature of the $X(3872)$ is still an open question. To explore its internal structure one step further, we study the effects of the Coulomb correction on the binding property for the $X(3872)$ and $[D^+D^{*-}]$ with $J^{PC}=1^{+-}$, $[D_s^+D^{*-}_s]$ with $J^{PC}=1^{++}$ and $[D_s^+D^{*-}_s]$ with $J^{PC}=1^{+-}$ in the hadronic molecular framework, and discuss their radiative decay behavior.

In addition to the $S$-$D$ wave mixing, isospin breaking and coupled channel effect, the electromagnetic correction is also included. For the $X(3872)$, the Coulomb interaction increases the binding energy by about 0.2$-$0.4 MeV with a cutoff parameter of about 1.17$-$1.19 GeV. The isospin breaking effect from the meson mass difference weakens the attraction between the composed hadrons in the $X(3872)$, while the Coulomb interaction strengthens it. The charge distribution has almost no effect on the binding solutions for the $X(3872)$, probably because the $X(3872)$ is a very loosely bound state.

Compared to the $D\bar{D}^*$ system, the Coulomb effect is more obvious for the $D_s\bar{D}_s^*$ system, which contains the purely charged channel. There are many molecules composed of charmed-strange mesons for the new hadron states, such as $[D_s^+D_s^-]$ for the $X(3960)$ \cite{Mutuk:2022ckn,Xin:2022bzt,Bayar:2022dqa,Prelovsek:2020eiw}, $[D_s^{*+}D_s^{*-}]$ for the $X(4140)$ or $X(4160)$ \cite{Liu:2009ei,Wang:2018djr}, $[D_s^{+}D_{s0}^{-}]$ for the X(4274) \cite{He:2013oma,Shen:2010ky,Finazzo:2011he}, $[D_s^{*+}D_{s0}^{*-}]$ for the $X(4350)$ \cite{Zhang:2009em}, and so on. We expect that the Coulomb interaction can play an important role in unravelling the nature of these hadrons. 

It is well known that the electromagnetic interaction is as important as the quark mass difference for the interpretation of many isospin breaking effects. However, the effect of the Coulomb interaction is relatively small in the bound solutions of the $D\bar{D}^*$ system, compared to the influence from the mass difference between charmed mesons. The reason is that the charged charmed meson is about 4 MeV heavier than the neutral one, and this is much larger than the neutron-proton mass difference. Thus the charged channel is relatively small in the $D\bar{D}^*$ system, which limits the role of the electromagnetic interaction in the bound solutions. This situation would change if the mass difference between the charm mesons in a hadronic molecule could become smaller than that in vacuum.   

It remains to be seen whether the neutral channel dominates in the $D\bar{D}^*$ system. The electromagnetic interaction explicitly breaks the isospin symmetry, and thus the electromagnetic decays encode more important information about the underlying structure of the hadronic molecules. The wave functions of the molecules are directly used in this work to study the radiative decays, which consequently reveal the structures of new hadronic states.

The $X(3872)$ decay to $\gamma\psi_n$ channels is a puzzle. We calculate the radiative decay width of the $X(3872)$ by assigning it as a $D\bar{D}^*$ hadronic molecular state. We can obtain the width of the $X(3872)\rightarrow \gamma J/\psi$ consistent with the experimental measurements. The experimental values for the ratio $ R_{\gamma\psi}$ are a little controversial, and our approach gives the $X(3872)\rightarrow \gamma\psi(2S)$ channel as suppressed and $ R_{\gamma\psi}\textless 1$, in agreement with the Belle and BES\Rmnum{3} measurements. We then predict the radiative width of the $[D\bar{D}^{*}]$ and $[D_s^+D^{*-}_s]$ molecules with $J^{PC}=1^{+-}$ decaying to $\gamma\eta_c$. The decay channel $\gamma\eta_c(2S)$ has a larger width than $\gamma\eta_c(1S)$. Although no experimental data on these radiative decay widths, we expect that the future experiments will be able to measure these decay channels in order to test our results with precise experimental data.

\section*{ACKNOWLEDGMENTS}
This work is supported by the National Natural Science Foundation of China under Grants No. 12175091, No.12335001, and No.12247101, the China National Funds for Distinguished Young Scientists under Grant No. 11825503, the National Key Research and Development Program of China under Contract No. 2020YFA0406400, the 111 Project under Grant No. B20063, the innovation project for young science and technology talents of Lanzhou city under Grant No. 2023-QN-107, the fundamental Research Funds for the Central Universities, and the project for top-notch innovative talents of Gansu province. F.L.W. is also supported by the China Postdoctoral Science Foundation under Grant No. 2022M721440. J.Z.W. is also supported by the National Postdoctoral Program for Innovative Talent.

\appendix
\section{THE DETAILS OF \linebreak POLARIZATION VECTORS}
The polarization vectors $\tilde\epsilon^\mu_{D^*}(\textbf{p},\lambda)$ with helicity $\lambda$ can be expressed by \cite{Zhu:1999ur}
\begin{equation}
\begin{split}
\tilde\epsilon^\mu_{D^*}(\textbf{p},\lambda=0)=&\left(\begin{array}{c}
|\textbf{p}|/m_{D^*}\\
\sin\theta \cos\phi \, E_{\rm \textbf{p}}/m_{D^*}\\
\sin\theta \sin\phi \, E_{\rm \textbf{p}}/m_{D^*}\\
\cos\theta~  E_{\rm \textbf{p}}/m_{D^*}
\end{array}
  \right),\\
  \end{split}
\end{equation}
and
\begin{equation}
\begin{split}
\tilde\epsilon^\mu_{D^*}(\textbf{p},\lambda=\pm1)=&\frac{1}{\sqrt{2}}\left(\begin{array}{c}
0\\
\mp \cos\theta\cos\phi+i\sin\phi\\
\mp \cos\theta\sin\phi-i \cos\phi\\
\pm \sin\theta
\end{array}
  \right).
  \end{split}
\end{equation}
We can obtain the polarization vectors of initial particle $\epsilon^\mu_{D^*,\,m_S}(\textbf{p})$ where $m_S$ is the z component of spin
\begin{eqnarray}
\epsilon^\mu_{D^*,\,m_S}(\textbf{p})&=&\sum_{\lambda}D^{j*}_{m_S \lambda}(\phi, \theta, 0)\,\tilde\epsilon^\mu_{D^*}(\textbf{p},\,\lambda). \label{zpolarization}
\end{eqnarray}

If $\textbf{k}$ is along the $z$ axis, the polarization vectors $\epsilon^{*\mu}_{\psi_n,m_S}$ of $\psi_n$ are
\begin{eqnarray}
\epsilon^{*\mu}_{\psi_n,0}(-\textbf{k})&=&\frac{1}{m_{\psi_n}}\left(|\textbf{k}|,0,0,-E_{\rm \textbf{k}}\right)^T,\nonumber\\
\epsilon^{*\mu}_{\psi_n,\pm1}(-\textbf{k})&=&\frac{1}{\sqrt{2}}(0,1,\pm i,0)^T,  \nonumber
\end{eqnarray}
and there exist only horizontal polarization vectors $\epsilon^{*\mu}_{\gamma,m_S}$ for photon which are $\epsilon^{*\mu}_{\gamma,\pm1}(\textbf{k})=\frac{1}{\sqrt{2}}(0,1,\mp i,0)$. For arbitrary directions of $\textbf{k}$, one can obtain the corresponding polarization vectors with the combination of above expressions and the Wigner $D$ functions.

\vfil



\begin{thebibliography}{200}
\bibitem{Belle:2003nnu}
S.~K.~Choi \textit{et al.} (Belle Collaboration),
Observation of a narrow charmonium-like state in exclusive $B^\pm \to K^\pm \pi^+ \pi^- J/\psi$ decays,
\href{https://journals.aps.org/prl/abstract/10.1103/PhysRevLett.91.262001}{Phys. Rev. Lett. \textbf{91}, 262001 (2003)}.

\bibitem{Godfrey:1985xj}
S.~Godfrey and N.~Isgur,
Mesons in a relativized quark model with chromodynamics,
\href{https://journals.aps.org/prd/abstract/10.1103/PhysRevD.32.189}{Phys. Rev. D \textbf{32}, 189 (1985)}.

\bibitem{Capstick:1986ter}
S.~Capstick and N.~Isgur,
Baryons in a relativized quark model with chromodynamics,
\href{https://journals.aps.org/prd/abstract/10.1103/PhysRevD.34.2809}{Phys. Rev. D \textbf{34}, 2809 (1986)}.

\bibitem{LHCb:2015yax}
R.~Aaij \textit{et al.} (LHCb Collaboration),
Observation of $J/\psi p$ resonances consistent with pentaquark states in $\Lambda_b^0 \to J/\psi K^- p$ decays,
\href{https://journals.aps.org/prl/abstract/10.1103/PhysRevLett.115.072001}{Phys. Rev. Lett. \textbf{115}, 072001 (2015)}.

\bibitem{LHCb:2019kea}
R.~Aaij \textit{et al.} (LHCb Collaboration),
Observation of a narrow pentaquark state, $P_c(4312)^+$, and of two-peak structure of the $P_c(4450)^+$,
\href{https://journals.aps.org/prl/abstract/10.1103/PhysRevLett.122.222001}{Phys. Rev. Lett. \textbf{122}, 222001 (2019)}.

\bibitem{LHCb:2020jpq}
R.~Aaij \textit{et al.} (LHCb Collaboration),
Evidence of a $J/\psi\Lambda$ structure and observation of excited $\Xi^-$ states in the $\Xi^-_b \to J/\psi\Lambda K^-$ decay,
\href{https://doi.org/10.1016/j.scib.2021.02.030}{Sci. Bull. \textbf{66}, 1278 (2021)}.

\bibitem{LHCb:2022ogu}
R.~Aaij \textit{et al.} (LHCb Collaboration),
Observation of a $J/\psi\Lambda$ resonance consistent with a strange pentaquark candidate in $B^- \rightarrow J/\psi \Lambda \bar{p}$ decays,
\href{https://journals.aps.org/prl/abstract/10.1103/PhysRevLett.131.031901}{Phys. Rev. Lett. \textbf{131}, 031901 (2023)}.

\bibitem{LHCb:2021vvq}
R.~Aaij \textit{et al.} (LHCb Collaboration),
Observation of an exotic narrow doubly charmed tetraquark,
\href{https://www.nature.com/articles/s41567-022-01614-y}{Nat. Phys. \textbf{18}, 751 (2022)}.

\bibitem{Swanson:2006st}
E.~S.~Swanson,
The new heavy mesons: a status report,
\href{https://doi.org/10.1016/j.physrep.2006.04.003}{Phys. Rep. \textbf{429}, 243 (2006)}.

\bibitem{Chen:2016qju}
H.~X.~Chen, W.~Chen, X.~Liu and S.~L.~Zhu,
The hidden-charm pentaquark and tetraquark states,
\href{https://doi.org/10.1016/j.physrep.2016.05.004}{Phys. Rep. \textbf{639}, 1 (2016)}.

\bibitem{Chen:2016spr}
H.~X.~Chen, W.~Chen, X.~Liu, Y.~R.~Liu and S.~L.~Zhu,
A review of the open charm and open bottom systems,
\href{https://iopscience.iop.org/article/10.1088/1361-6633/aa6420}{Rep. Prog. Phys. \textbf{80}, 076201 (2017)}.

\bibitem{Esposito:2016noz}
A.~Esposito, A.~Pilloni and A.~D.~Polosa,
Multiquark resonances,
\href{https://doi.org/10.1016/j.physrep.2016.11.002}{Phys. Rep. \textbf{668}, 1 (2017)}.

\bibitem{Olsen:2017bmm}
S.~L.~Olsen, T.~Skwarnicki and D.~Zieminska,
Nonstandard heavy mesons and baryons: Experimental evidence,
\href{https://journals.aps.org/rmp/pdf/10.1103/RevModPhys.90.015003}{Rev. Mod. Phys. \textbf{90}, 015003 (2018)}.

\bibitem{Brambilla:2019esw}
N.~Brambilla, S.~Eidelman, C.~Hanhart, A.~Nefediev, C.~P.~Shen, C.~E.~Thomas, A.~Vairo and C.~Z.~Yuan,
The $XYZ$ states: Experimental and theoretical status and perspectives,
\href{https://doi.org/10.1016/j.physrep.2020.05.001}{Phys. Rep. \textbf{873}, 1 (2020)}.

\bibitem{Liu:2019zoy}
Y.~R.~Liu, H.~X.~Chen, W.~Chen, X.~Liu and S.~L.~Zhu,
Pentaquark and Tetraquark states,
\href{https://doi.org/10.1016/j.ppnp.2019.04.003}{Prog. Part. Nucl. Phys. \textbf{107}, 237 (2019)}.

\bibitem{Chen:2022asf}
H.~X.~Chen, W.~Chen, X.~Liu, Y.~R.~Liu and S.~L.~Zhu,
An updated review of the new hadron states,
\href{https://iopscience.iop.org/article/10.1088/1361-6633/aca3b6}{Rept. Prog. Phys. \textbf{86}, 026201 (2023)}.

\bibitem{Swanson:2003tb}
E.~S.~Swanson,
Short range structure in the $X(3872)$,
\href{https://doi.org/10.1016/j.physletb.2004.03.033}{Phys. Lett. B \textbf{588}, 189 (2004)}.

\bibitem{Wong:2003xk}
C.~Y.~Wong,
Molecular states of heavy quark mesons,
\href{https://journals.aps.org/prc/abstract/10.1103/PhysRevC.69.055202}{Phys. Rev. C \textbf{69}, 055202 (2004)}.

\bibitem{Close:2003sg}
F.~E.~Close and P.~R.~Page,
The $D^{*0}$ anti-$D^0$ threshold resonance,
\href{https://doi.org/10.1016/j.physletb.2003.10.032}{Phys. Lett. B \textbf{578}, 119 (2004)}.

\bibitem{Voloshin:2003nt}
M.~B.~Voloshin,
Interference and binding effects in decays of possible molecular component of $X(3872)$,
\href{https://doi.org/10.1016/j.physletb.2003.11.014}{Phys. Lett. B \textbf{579}, 316 (2004)}.


\bibitem{Liu:2008fh}
Y.~R.~Liu, X.~Liu, W.~Z.~Deng and S.~L.~Zhu,
Is $X(3872)$ really a molecular state?,
\href{https://link.springer.com/article/10.1140/epjc/s10052-008-0640-4}{Eur. Phys. J. C \textbf{56}, 63 (2008)}.

\bibitem{Liu:2009qhy}
X.~Liu, Z.~G.~Luo, Y.~R.~Liu, and S.~L.~Zhu,
$X(3872)$ and other possible heavy molecular states,
\href{https://link.springer.com/article/10.1140/epjc/s10052-009-1020-4}{Eur. Phys. J. C \textbf{61}, 411 (2009)}.

\bibitem{AlFiky:2005jd}
M.~T.~AlFiky, F.~Gabbiani and A.~A.~Petrov,
$X(3872)$: Hadronic molecules in effective field theory,
\href{https://doi.org/10.1016/j.physletb.2006.07.069}{Phys. Lett. B \textbf{640}, 238 (2006)}.

\bibitem{Thomas:2008ja}
C.~E.~Thomas and F.~E.~Close,
Is $X(3872)$ a molecule?,
\href{https://journals.aps.org/prd/abstract/10.1103/PhysRevD.78.034007}{Phys. Rev. D \textbf{78}, 034007 (2008)}.

\bibitem{Tornqvist:1993vu}
N.~A.~T\"{o}rnqvist,
On deusons or deuteron-like meson meson bound states,
\href{https://link.springer.com/article/10.1007/BF02734018}{Nuovo Cimento Soc. Ital. Fis. \textbf{107A}, 2471 (1994)}.

\bibitem{Tornqvist:2004qy}
N.~A.~Tornqvist,
Isospin breaking of the narrow charmonium state of Belle at 3872-MeV as a deuson,
\href{https://doi.org/10.1016/j.physletb.2004.03.077}{Phys. Lett. B \textbf{590}, 209 (2004)}.

\bibitem{Lee:2009hy}
I.~W.~Lee, A.~Faessler, T.~Gutsche, and V.~E.~Lyubovitskij,
$X(3872)$ as a molecular $D\bar{D}^*$ state in a potential model,
\href{https://journals.aps.org/prd/abstract/10.1103/PhysRevD.80.094005}{Phys. Rev. D \textbf{80}, 094005 (2009)}.

\bibitem{Braaten:2010mg}
E.~Braaten, H.~W.~Hammer, and T.~Mehen,
Scattering of an ultrasoft Pion and the $X(3872)$,
\href{https://journals.aps.org/prd/abstract/10.1103/PhysRevD.82.034018}{Phys. Rev. D \textbf{82}, 034018 (2010)}.

\bibitem{Wang:2013kva}
P.~Wang and X.~G.~Wang,
Study on $X(3872)$ from effective field theory with pion exchange interaction,
\href{https://journals.aps.org/prl/abstract/10.1103/PhysRevLett.111.042002}{Phys. Rev. Lett. \textbf{111}, 042002 (2013)}.

\bibitem{Baru:2013rta}
V.~Baru, E.~Epelbaum, A.~A.~Filin, C.~Hanhart, U.~G.~Meissner, and A.~V.~Nefediev,
Quark mass dependence of the $X(3872)$ binding energy,
\href{https://doi.org/10.1016/j.physletb.2013.08.073}{Phys. Lett. B \textbf{726}, 537 (2013)}.

\bibitem{Baru:2015nea}
V.~Baru, E.~Epelbaum, A.~A.~Filin, F.~K.~Guo, H.~W.~Hammer, C.~Hanhart, U.~G.~Mei\ss{}ner, and A.~V.~Nefediev,
Remarks on study of $X(3872)$ from effective field theory with pion-exchange interaction,
\href{https://journals.aps.org/prd/abstract/10.1103/PhysRevD.91.034002}{Phys. Rev. D \textbf{91}, 034002 (2015)}.

\bibitem{Song:2023pdq}
J.~Song, L.~R.~Dai, and E.~Oset,
Evolution of compact states to molecular ones with coupled channels: The case of the $X(3872)$,
\href{https://doi.org/10.1103/PhysRevD.108.114017}{Phys. Rev. D \textbf{108}, 114017 (2023)}.

\bibitem{Gamermann:2009uq}
D.~Gamermann, J.~Nieves, E.~Oset, and E.~Ruiz Arriola,
Couplings in coupled channels versus wave functions: application to the $X(3872)$ resonance,
\href{https://doi.org/10.1103/PhysRevD.81.014029}{Phys. Rev. D \textbf{81}, 014029 (2010)}.

\bibitem{Maiani:2004vq}
L.~Maiani, F.~Piccinini, A.~D.~Polosa, and V.~Riquer,
Diquark-antidiquarks with hidden or open charm and the nature of $X(3872)$,
\href{https://journals.aps.org/prd/abstract/10.1103/PhysRevD.71.014028}{Phys. Rev. D \textbf{71}, 014028 (2005)}.

\bibitem{Ebert:2005nc}
D.~Ebert, R.~N.~Faustov, and V.~O.~Galkin,
Masses of heavy tetraquarks in the relativistic quark model,
\href{https://doi.org/10.1016/j.physletb.2006.01.026}{Phys. Lett. B \textbf{634}, 214 (2006)}.

\bibitem{Hogaasen:2005jv}
H.~Hogaasen, J.~M.~Richard, and P.~Sorba,
A chromomagnetic mechanism for the $X(3872)$ resonance,
\href{https://journals.aps.org/prd/abstract/10.1103/PhysRevD.73.054013}{Phys. Rev. D \textbf{73}, 054013 (2006)}.

\bibitem{Barnea:2006sd}
N.~Barnea, J.~Vijande, and A.~Valcarce,
Four-quark spectroscopy within the hyperspherical formalism,
\href{https://journals.aps.org/prd/abstract/10.1103/PhysRevD.73.054004}{Phys. Rev. D \textbf{73}, 054004 (2006)}.

\bibitem{Matheus:2006xi}
R.~D.~Matheus, S.~Narison, M.~Nielsen, and J.~M.~Richard,
Can the $X(3872)$ be a $1^{++}$ four-quark state?,
\href{https://journals.aps.org/prd/abstract/10.1103/PhysRevD.75.014005}{Phys. Rev. D \textbf{75}, 014005 (2007)}.

\bibitem{Vijande:2007fc}
J.~Vijande, E.~Weissman, N.~Barnea, and A.~Valcarce,
Do $c\bar{c}n\bar{n}$ bound states exist?,
\href{https://journals.aps.org/prd/abstract/10.1103/PhysRevD.76.094022}{Phys. Rev. D \textbf{76}, 094022 (2007)}.

\bibitem{Wang:2013vex}
Z.~G.~Wang and T.~Huang,
Analysis of the $X(3872)$, $Z_c(3900)$ and $Z_c(3885)$ as axial-vector tetraquark states with QCD sum rules,
\href{https://journals.aps.org/prd/abstract/10.1103/PhysRevD.89.054019}{Phys. Rev. D \textbf{89}, 054019 (2014)}.

\bibitem{Eichten:2004uh}
E.~J.~Eichten, K.~Lane, and C.~Quigg,
Charmonium levels near threshold and the narrow state $X(3872) \rightarrow \pi^{+}\pi^{-}J/\psi$,
\href{https://journals.aps.org/prd/abstract/10.1103/PhysRevD.69.094019}{Phys. Rev. D \textbf{69}, 094019 (2004)}.

\bibitem{Barnes:2003vb}
T.~Barnes and S.~Godfrey,
Charmonium options for the $X(3872)$,
\href{https://journals.aps.org/prd/abstract/10.1103/PhysRevD.69.054008}{Phys. Rev. D \textbf{69}, 054008 (2004)}.


\bibitem{Suzuki:2005ha}
M.~Suzuki,
The $X(3872)$ boson: Molecule or charmonium,
\href{https://journals.aps.org/prd/abstract/10.1103/PhysRevD.72.114013}{Phys. Rev. D \textbf{72}, 114013 (2005)}.

\bibitem{Kong:2006ni}
Y.~M.~Kong and A.~Zhang,
Charmonium possibility of $X(3872)$,
\href{https://doi.org/10.1016/j.physletb.2007.09.062}{Phys. Lett. B \textbf{657}, 192 (2007)}.

\bibitem{Voloshin:2007dx}
M.~B.~Voloshin,
Charmonium,
\href{https://doi.org/10.1016/j.ppnp.2008.02.001}{Prog. Part. Nucl. Phys. \textbf{61}, 455 (2008)}.

\bibitem{Kalashnikova:2010hv}
Y.~S.~Kalashnikova and A.~V.~Nefediev,
$X(3872)$ as a $^1D_2$ charmonium state,
\href{https://journals.aps.org/prd/abstract/10.1103/PhysRevD.82.097502}{Phys. Rev. D \textbf{82}, 097502 (2010)}.

\bibitem{Ferretti:2013faa}
J.~Ferretti, G.~Galat\`a, and E.~Santopinto,
Interpretation of the $X(3872)$ as a charmonium state plus an extra component due to the coupling to the meson-meson continuum,
\href{https://journals.aps.org/prc/abstract/10.1103/PhysRevC.88.015207}{Phys. Rev. C \textbf{88}, 015207 (2013)}.

\bibitem{Ferretti:2014xqa}
J.~Ferretti, G.~Galat\`a, and E.~Santopinto,
Quark structure of the $X(3872)$ and $\chi_b(3P)$ resonances,
\href{https://journals.aps.org/prd/abstract/10.1103/PhysRevD.90.054010}{Phys. Rev. D \textbf{90}, 054010 (2014)}.

\bibitem{Li:2004sta}
B.~A.~Li,
Is $X(3872)$ a possible candidate of hybrid meson,
\href{https://doi.org/10.1016/j.physletb.2004.11.062}{Phys. Lett. B \textbf{605}, 306 (2005)}.

\bibitem{Meng:2005er}
C.~Meng, Y.~J.~Gao, and K.~T.~Chao,
$B \rightarrow \chi_{c1}(1P,2P)K$ decays in QCD factorization and $X(3872)$,
\href{https://journals.aps.org/prd/abstract/10.1103/PhysRevD.87.074035}{Phys. Rev. D \textbf{87}, 074035 (2013)}.

\bibitem{Meng:2014ota}
C.~Meng, J.~J.~Sanz-Cillero, M.~Shi, D.~L.~Yao, and H.~Q.~Zheng,
Refined analysis on the $X(3872)$ resonance,
\href{https://journals.aps.org/prd/abstract/10.1103/PhysRevD.92.034020}{Phys. Rev. D \textbf{92}, 034020 (2015)}.


\bibitem{ParticleDataGroup:2022pth}
R.~L.~Workman \textit{et al.} (Particle Data Group),
Review of particle physics,
\href{https://doi.org/10.1093/ptep/ptac097}{Prog. Theor. Exp. Phys. \textbf{2022}, 083C01 (2022)}.

\bibitem{Baru:2016iwj}
V.~Baru, E.~Epelbaum, A.~A.~Filin, C.~Hanhart, U.~G.~Mei\ss{}ner, and A.~V.~Nefediev,
Heavy-quark spin symmetry partners of the $X (3872)$ revisited,
\href{https://doi.org/10.1016/j.physletb.2016.10.008}{Phys. Lett. B \textbf{763}, 20 (2016)}.

\bibitem{COMPASS:2017wql}
M.~Aghasyan \textit{et al.} (COMPASS Collaboration),
Search for muoproduction of $X(3872)$ at COMPASS and indication of a new state $\tilde{X}(3872)$,
\href{https://doi.org/10.1016/j.physletb.2018.07.008}{Phys. Lett. B \textbf{783}, 334 (2018)}.

\bibitem{Gamermann:2007mu}
D.~Gamermann and E.~Oset,
Hidden charm dynamically generated resonances and the $e^+ e^- \to J/\psi D\bar{D}, J/\psi D\bar{D}^*$ reactions,
\href{https://doi.org/10.1140/epja/i2007-10580-5}{Eur. Phys. J. A \textbf{36}, 189 (2008)}.

\bibitem{Dai:2020yfu}
L.~Dai, G.~Toledo, and E.~Oset,
Searching for a $D\bar{D}$ bound state with the $\psi(3770) \to \gamma D^0 \bar{D}^0$ decay,
\href{https://doi.org/10.1140/epjc/s10052-020-8058-8}{Eur. Phys. J. C \textbf{80}, 510 (2020)}.

\bibitem{Wang:2020elp}
E.~Wang, H.~S.~Li, W.~H.~Liang, and E.~Oset,
Analysis of the $\gamma\gamma \to D\bar{D}$ reaction and the $D\bar{D}$ bound state,
\href{https://doi.org/10.1103/PhysRevD.103.054008}{Phys. Rev. D \textbf{103}, 054008 (2021)}.

\bibitem{Belle:2005rte}
S.~Uehara \textit{et al.} (Belle Collaboration),
Observation of a $\chi_{c2}^\prime$ candidate in $\gamma \gamma \to D\bar{D}$ production at BELLE,
\href{https://doi.org/10.1103/PhysRevLett.96.082003}{Phys. Rev. Lett. \textbf{96}, 082003 (2006)}.

\bibitem{Belle:2007woe}
P.~Pakhlov \textit{et al.} (Belle Collaboration),
Production of new charmoniumlike States in $e^+e^-\to J/\psi D^{(*)}\bar{D}^{(*)}$ at $\sqrt{s} \approx 10.6$ GeV,
\href{https://doi.org/10.1103/PhysRevLett.100.202001}{Phys. Rev. Lett. \textbf{100}, 202001 (2008)}.

\bibitem{BaBar:2010jfn}
B.~Aubert \textit{et al.} (BaBar Collaboration),
Observation of the $\chi_{c2}(2p)$ meson in the reaction $\gamma \gamma \to D \bar{D}$ at \textit{BABAR},
\href{https://doi.org/10.1103/PhysRevD.81.092003}{Phys. Rev. D \textbf{81}, 092003 (2010)}.

\bibitem{Cincioglu:2016fkm}
E.~Cincioglu, J.~Nieves, A.~Ozpineci, and A.~U.~Yilmazer,
Quarkonium contribution to meson molecules,
\href{https://doi.org/10.1140/epjc/s10052-016-4413-1}{Eur. Phys. J. C \textbf{76}, 576 (2016)}.

\bibitem{Ortega:2020tng}
P.~G.~Ortega and D.~R.~Entem,
Coupling hadron-hadron thresholds within a chiral quark model approach,
\href{https://doi.org/10.3390/sym13020279}{Symmetry \textbf{13}, 279 (2021)}.

\bibitem{Ortega:2017qmg}
P.~G.~Ortega, J.~Segovia, D.~R.~Entem, and F.~Fern\'andez,
Charmonium resonances in the 3.9 GeV/$c^2$ energy region and the $X(3915)/X(3930)$ puzzle,
\href{https://doi.org/10.1016/j.physletb.2018.01.005}{Phys. Lett. B \textbf{778}, 1 (2018)}.



\bibitem{Li:2012cs}
N.~Li and S.~L.~Zhu,
Isospin breaking, coupled-channel effects and diagnosis of $X(3872)$,
\href{https://journals.aps.org/prd/abstract/10.1103/PhysRevD.86.074022}{Phys. Rev. D \textbf{86}, 074022 (2012)}.





\bibitem{Gamermann:2009fv}
D.~Gamermann and E.~Oset,
Isospin breaking effects in the $X(3872)$ resonance,
\href{https://journals.aps.org/prd/abstract/10.1103/PhysRevD.80.014003}{Phys. Rev. D \textbf{80}, 014003 (2009)}.

\bibitem{Takizawa:2010rxa}
M.~Takizawa and S.~Takeuchi,
Structure of the $X(3872)$ and its isospin symmetry breaking,
\href{https://doi.org/10.1051/epjconf/20100303026}{EPJ Web Conf. \textbf{3}, 03026 (2010)}.

\bibitem{Zhao:2014gqa}
L.~Zhao, L.~Ma, and S.~L.~Zhu,
Spin-orbit force, recoil corrections, and possible $B \bar{B}^{*}$ and $D \bar{D}^{*}$  molecular states,
\href{https://journals.aps.org/prd/abstract/10.1103/PhysRevD.89.094026}{Phys. Rev. D \textbf{89}, 094026 (2014)}.

\bibitem{Lyu:2021qsh}
Y.~Lyu, H.~Tong, T.~Sugiura, S.~Aoki, T.~Doi, T.~Hatsuda, J.~Meng and T.~Miyamoto,
Dibaryon with highest charm number near unitarity from lattice QCD,
\href{https://journals.aps.org/prl/abstract/10.1103/PhysRevLett.127.072003}{Phys. Rev. Lett. \textbf{127}, 072003 (2021)}.

\bibitem{Liu:2021pdu}
M.~Z.~Liu and L.~S.~Geng,
Prediction of an $\Omega_{bbb}\Omega_{bbb}$ dibaryon in the extended one-boson exchange model,
\href{https://iopscience.iop.org/article/10.1088/0256-307X/38/10/101201}{Chin. Phys. Lett. \textbf{38}, 101201 (2021)}.

\bibitem{Mathur:2022nez}
N.~Mathur, M.~Padmanath and D.~Chakraborty,
Strongly bound dibaryon with maximal beauty flavor from lattice QCD,
\href{https://doi.org/10.1103/PhysRevLett.130.111901}{Phys.Rev.Lett. 130 111901 (2023)}.

\bibitem{Wu:2021kbu}
T.~W.~Wu, Y.~W.~Pan, M.~Z.~Liu, S.~Q.~Luo, L.~S.~Geng and X.~Liu,
Discovery of the doubly charmed $T_{cc}^+$ state implies a triply charmed $H_{ccc}$ hexaquark state,
\href{https://journals.aps.org/prd/abstract/10.1103/PhysRevD.105.L031505}{Phys. Rev. D \textbf{105}, L031505 (2022)}.

\bibitem{Zhang:2020mpi}
Z.~H.~Zhang and F.~K.~Guo,
$D^{\pm}D^*{\mp}$ hadronic atom as a Key to revealing the $X(3872)$ mystery,
\href{https://journals.aps.org/prl/abstract/10.1103/PhysRevLett.127.012002}{Phys. Rev. Lett. \textbf{127}, 012002 (2021)}.

\bibitem{Swanson:2004pp}
E.~S.~Swanson,
Diagnostic decays of the $X(3872)$,
\href{https://doi.org/10.1016/j.physletb.2004.07.059}{Phys. Lett. B \textbf{598}, 197 (2004)}.

\bibitem{Guo:2014taa}
F.~K.~Guo, C.~Hanhart, Y.~S.~Kalashnikova, U.~G.~Mei\ss{}ner and A.~V.~Nefediev,
What can radiative decays of the $X(3872)$ teach us about its nature?,
\href{https://doi.org/10.1016/j.physletb.2015.02.013}{Phys. Lett. B \textbf{742}, 394 (2015)}.

\bibitem{Badalian:2012jz}
A.~M.~Badalian, V.~D.~Orlovsky, Y.~A.~Simonov, and B.~L.~G.~Bakker,
The ratio of decay widths of $X(3872)$ to $\psi^{\prime}\gamma $ and $J/\psi\gamma$ as a test of the $X(3872)$ dynamical structure,
\href{https://journals.aps.org/prd/abstract/10.1103/PhysRevD.85.114002}{Phys. Rev. D \textbf{85}, 114002 (2012)}.

\bibitem{Wang:2010ej}
T.~H.~Wang and G.~L.~Wang,
Radiative E1 decays of $X(3872)$,
\href{https://doi.org/10.1016/j.physletb.2011.02.014}{Phys. Lett. B \textbf{697}, 233 (2011)}.

\bibitem{Mehen:2011ds}
T.~Mehen and R.~Springer,
Radiative decays $X(3872) \to \psi(2S)\gamma$ and $\psi(4040)$ $\rightarrow$ $X(3872)\gamma$ in effective field theory,
\href{https://journals.aps.org/prd/abstract/10.1103/PhysRevD.83.094009}{Phys. Rev. D \textbf{83}, 094009 (2011)}.

\bibitem{Li:2009zu}
B.~Q.~Li and K.~T.~Chao,
Higher charmonia and X,Y,Z states with screened potential,
\href{https://journals.aps.org/prd/abstract/10.1103/PhysRevD.79.094004}{Phys. Rev. D \textbf{79}, 094004 (2009)}.

\bibitem{Lahde:2002wj}
T.~A.~Lahde,
Exchange current operators and electromagnetic dipole transitions in heavy quarkonia,
\href{https://doi.org/10.1016/S0375-9474(02)01362-3}{Nucl. Phys. A \textbf{714}, 183 (2003)}.

\bibitem{Barnes:2005pb}
T.~Barnes, S.~Godfrey, and E.~S.~Swanson,
Higher charmonia,
\href{https://journals.aps.org/prd/abstract/10.1103/PhysRevD.72.054026}{Phys. Rev. D \textbf{72}, 054026 (2005)}.

\bibitem{Yu:2023nxk}
S.~Y.~Yu and X.~W.~Kang,
Nature of $X(3872)$ from its radiative decay,
\href{https://doi.org/10.1016/j.physletb.2023.138404}{Phys. Lett. B \textbf{848}, 138404 (2024)}.

\bibitem{Dong:2009uf}
Y.~Dong, A.~Faessler, T.~Gutsche, and V.~E.~Lyubovitskij,
$J/\psi\gamma$ and $\psi(2S)\gamma$ decay modes of the $X(3872)$,
\href{https://iopscience.iop.org/article/10.1088/0954-3899/38/1/015001}{J. Phys. G \textbf{38}, 015001 (2011)}.

\bibitem{Eichten:2005ga}
E.~J.~Eichten, K.~Lane and C.~Quigg,
New states above charm threshold,
\href{https://journals.aps.org/prd/abstract/10.1103/PhysRevD.73.014014}{Phys. Rev. D \textbf{73}, 014014 (2006);}
\href{https://journals.aps.org/prd/abstract/10.1103/PhysRevD.73.014014}{\textbf{73}, 079903(E) (2006)]}.

\bibitem{Cardoso:2014xda}
M.~Cardoso, G.~Rupp, and E.~van Beveren,
Unquenched quark-model calculation of $X(3872)$ electromagnetic decays,
\href{https://link.springer.com/article/10.1140/epjc/s10052-014-3254-z}{Eur. Phys. J. C \textbf{75}, 26 (2015)}.



\bibitem{Meng:2020cbk}
L.~Meng, B.~Wang, and S.~L.~Zhu,
Predicting the $\bar{D}^{(*)}_sD_s^{(*)}$ bound states as the partners of $X(3872)$,
\href{https://doi.org/10.1016/j.scib.2021.03.016}{Sci. Bull. \textbf{66}, 1288 (2021)}.

\bibitem{Shi:2023mer}
P.~P.~Shi, J.~M.~Dias, and F.~K.~Guo,
Radiative decays of the spin-2 partner of $X(3872)$,
\href{https://doi.org/10.1016/j.physletb.2023.137987}{Phys. Lett. B \textbf{843}, 137987 (2023)}.

\bibitem{Belle:2011wdj}
V.~Bhardwaj \textit{et al.} (Belle Collaboration),
Observation of $X(3872)\to J/\psi \gamma$ and search for $X(3872)\to\psi'\gamma$ in B decays,
\href{https://journals.aps.org/prl/abstract/10.1103/PhysRevLett.107.091803}{Phys. Rev. Lett. \textbf{107}, 091803 (2011)}.

\bibitem{LHCb:2013kgk}
R.~Aaij \textit{et al.} (LHCb Collaboration),
Determination of the $X(3872)$ meson quantum numbers,
\href{https://journals.aps.org/prl/abstract/10.1103/PhysRevLett.110.222001}{Phys. Rev. Lett. \textbf{110}, 222001 (2013)}.


\bibitem{Grinstein:1992qt}
B.~Grinstein, E.~E.~Jenkins, A.~V.~Manohar, M.~J.~Savage, and M.~B.~Wise,
Chiral perturbation theory for $f_{D_{(s)}}$/$f_D$ and $B_{B_{(s)}}$/$B_B$,
\href{https://doi.org/10.1016/0550-3213(92)90248-A}{Nucl. Phys. B \textbf{380}, 369 (1992)}.


\bibitem{Casalbuoni:1992gi}
R.~Casalbuoni, A.~Deandrea, N.~Di Bartolomeo, R.~Gatto, F.~Feruglio, and G.~Nardulli,
Light vector resonances in the effective chiral Lagrangian for heavy mesons,
\href{https://doi.org/10.1016/0370-2693(92)91189-G}{Phys. Lett. B \textbf{292}, 371 (1992)}.

\bibitem{Casalbuoni:1996pg}
R.~Casalbuoni, A.~Deandrea, N.~Di Bartolomeo, R.~Gatto, F.~Feruglio, and G.~Nardulli,
Phenomenology of heavy meson chiral Lagrangians,
\href{https://doi.org/10.1016/S0370-1573(96)00027-0}{Phys. Rep. \textbf{281}, 145 (1997)}.

\bibitem{Yan:1992gz}
T.~M.~Yan, H.~Y.~Cheng, C.~Y.~Cheung, G.~L.~Lin, Y.~C.~Lin, and H.~L.~Yu,
Heavy quark symmetry and chiral dynamics,
\href{https://journals.aps.org/prd/abstract/10.1103/PhysRevD.46.1148}{Phys. Rev. D \textbf{46}, 1148 (1992); \textbf{55} 5851(E) (1997)}.

\bibitem{Bando:1987br}
M.~Bando, T.~Kugo, and K.~Yamawaki,
Nonlinear Realization and Hidden Local Symmetries,
\href{https://doi.org/10.1016/0370-1573(88)90019-1}{Phys. Rep. \textbf{164}, 217 (1988)}.

\bibitem{Harada:2003jx}
M.~Harada and K.~Yamawaki,
Hidden local symmetry at loop: A new perspective of composite gauge boson and chiral phase transition,
\href{https://doi.org/10.1016/S0370-1573(03)00139-X}{Phys. Rep. \textbf{381}, 1 (2003)}.

\bibitem{Ding:2008gr}
G.~J.~Ding,
Are $Y(4260)$ and $Z_2^+(4250)$ are $D_{(1)}D$ or $D_{(0)}D^*$ hadronic molecules?,
\href{https://journals.aps.org/prd/abstract/10.1103/PhysRevD.79.014001}{Phys. Rev. D \textbf{79}, 014001 (2009)}.

\bibitem{Sun:2011uh}
Z.~F.~Sun, J.~He, X.~Liu, Z.~G.~Luo and S.~L.~Zhu,
$Z_b(10610)^\pm$ and $Z_b(10650)^\pm$ as the $B^*\bar{B}$ and $B^*\bar{B}^{*}$ molecular states,
\href{https://journals.aps.org/prd/abstract/10.1103/PhysRevD.84.054002}{Phys. Rev. D \textbf{84}, 054002 (2011)}.

\bibitem{Wang:2020dya}
F.~L.~Wang and X.~Liu,
Exotic double-charm molecular states with hidden or open strangeness and around $4.5\sim 4.7$ GeV,
\href{https://journals.aps.org/prd/abstract/10.1103/PhysRevD.102.094006}{Phys. Rev. D \textbf{102}, 094006 (2020)}.

\bibitem{Berestetsky:1982}
V. B. Berestetsky, E. M. Lifshitz, and L. P. Pitaevsky, \textit{Quantum Electrodynamics} (Pergamon Press, New York,
1982).

\bibitem{Wang:2021ajy}
F.~L.~Wang, R.~Chen, and X.~Liu,
A new group of doubly charmed molecule with T-doublet charmed meson pair,
\href{https://doi.org/10.1016/j.physletb.2022.137502}{Phys. Lett. B \textbf{835}, 137502 (2022)}.

\bibitem{Chen:2017jjn}
R.~Chen, A.~Hosaka, and X.~Liu,
Prediction of triple-charm molecular pentaquarks,
\href{https://journals.aps.org/prd/abstract/10.1103/PhysRevD.96.114030}{Phys. Rev. D \textbf{96}, 114030 (2017)}.

\bibitem{Wang:2019nwt}
F.~L.~Wang, R.~Chen, Z.~W.~Liu, and X.~Liu,
Probing new types of $P_c$ states inspired by the interaction between $S$-wave charmed baryon and anti-charmed meson in a $\bar T$ doublet,
\href{https://journals.aps.org/prc/abstract/10.1103/PhysRevC.101.025201}{Phys. Rev. C \textbf{101}, 025201 (2020)}.

\bibitem{Tornqvist:1993ng}
N.~A.~T\"{o}rnqvist,
From the deuteron to deusons, an analysis of deuteronlike meson-meson bound states,
\href{https://link.springer.com/article/10.1007/BF01413192}{Z. Phys. C \textbf{61}, 525 (1994)}.

\bibitem{Liu:2007bf}
X.~Liu, Y.~R.~Liu, W.~Z.~Deng, and S.~L.~Zhu,
Is $Z^+(4430)$ a loosely bound molecular state?,
\href{https://journals.aps.org/prd/pdf/10.1103/PhysRevD.77.034003}{Phys. Rev. D \textbf{77}, 034003 (2008)}.

\bibitem{Cheng:2022qcm}
J.~B.~Cheng, Z.~Y.~Lin, and S.~L.~Zhu,
Double-charm tetraquark under the complex scaling method,
\href{https://doi.org/10.1103/PhysRevD.106.016012}{Phys. Rev. D \textbf{106}, 016012 (2022)}.


\bibitem{Schmidt:2018vvl}
M.~Schmidt, M.~Jansen, and H.~W.~Hammer,
Threshold effects and the line shape of the $X(3872)$ in effective field theory,
\href{https://doi.org/10.1103/PhysRevD.98.014032}{Phys. Rev. D \textbf{98}, 014032 (2018)}.


\bibitem{Baru:2011rs}  
V.~Baru, A.~A.~Filin, C.~Hanhart, Y.~S.~Kalashnikova, A.~E.~Kudryavtsev, and A.~V.~Nefediev,
Three-body $D\bar{D}\pi$ dynamics for the $X(3872)$,
\href{https://doi.org/10.1103/PhysRevD.84.074029}{Phys. Rev. D \textbf{84}, 074029 (2011)}.

\bibitem{Abrashkevich:1998nm}
A.~Abrashkevich, D.~Abrashkevich, M.~Kaschiev, and I.~Puzynin,
FESSDE 2.2: A new version of a program for the finite-element solution of the coupled-channel Schr\"{o}dinger equation using high-order accuracy approximations,
\href{https://doi.org/10.1016/S0010-4655(98)00099-X}{Comput. Phys. Commun. \textbf{115}, 90 (1998)}.

\bibitem{Abrashkevich:1995tb}
A.~Abrashkevich, D.~Abrashkevich, M.~Kaschiev, and I.~Puzynin,
FESSDE, a program for the finite-element solution of the coupled-channel Schrdinger equation using high-order accuracy approximations,
\href{https://doi.org/10.1016/0010-4655(94)00107-D}{Comput. Phys. Commun. \textbf{85}, 65 (1995)}.

\bibitem{Isola:2003fh}
C.~Isola, M.~Ladisa, G.~Nardulli and P.~Santorelli,
Charming penguins in $B\to K^*\pi$, $K(\rho,\omega,\phi)$ decays,
\href{https://journals.aps.org/prd/abstract/10.1103/PhysRevD.68.114001}{Phys. Rev. D \textbf{68}, 114001 (2003)}.

\bibitem{Liu:2008xz}
X.~Liu, Y.~R.~Liu, W.~Z.~Deng, and S.~L.~Zhu,
$Z^+(4430)$ as a $D_1^\prime D^* (D_1D^*)$ molecular state,
\href{https://journals.aps.org/prd/abstract/10.1103/PhysRevD.77.094015}{Phys. Rev. D \textbf{77}, 094015 (2008)}.

\bibitem{Chen:2017vai}
R.~Chen, A.~Hosaka, and X.~Liu,
Heavy molecules and one-$\sigma/\omega$-exchange model,
\href{https://doi.org/10.1103/PhysRevD.96.116012}{Phys. Rev. D \textbf{96}, 116012 (2017)}.

\bibitem{Wiringa:1994wb}
R.~B.~Wiringa, V.~G.~J.~Stoks, and R.~Schiavilla,
An accurate nucleon-nucleon potential with charge independence breaking,
\href{https://journals.aps.org/prc/abstract/10.1103/PhysRevC.51.38}{Phys. Rev. C \textbf{51}, 38 (1995)}.

\bibitem{Can:2012tx}
K.~U.~Can, G.~Erkol, M.~Oka, A.~Ozpineci, and T.~T.~Takahashi,
Vector and axial-vector couplings of $D$ and $D^*$ mesons in $2+1$ flavor lattice QCD,
\href{https://doi.org/10.1016/j.physletb.2012.12.050}{Phys. Lett. B \textbf{719}, 103 (2013)}.

\bibitem{Luo:2023hnp}
S.~Q.~Luo, Z.~W.~Liu, and X.~Liu,
New type of hydrogenlike charm-pion or charm-kaon matter,
\href{https://doi.org/10.1103/PhysRevD.107.054022}{Phys. Rev. D \textbf{107}, 054022 (2023)}.

\bibitem{Zhang:2006ix}
Y.~J.~Zhang, H.~C.~Chiang, P.~N.~Shen, and B.~S.~Zou,
Possible $S$-wave bound-states of two pseudoscalar mesons,
\href{https://doi.org/10.1103/PhysRevD.74.014013}{Phys. Rev. D \textbf{74}, 014013 (2006)}.

\bibitem{Colangelo:2003sa}
P.~Colangelo, F.~De Fazio, and T.~N.~Pham,
Nonfactorizable contributions in $B$ decays to charmonium: The case of $B^-\rightarrow K^- h_c$,
\href{https://journals.aps.org/prd/abstract/10.1103/PhysRevD.69.054023}{Phys. Rev. D \textbf{69}, 054023 (2004)}.

\bibitem{Lin:1999ad}
Z.~W.~Lin and C.~M.~Ko,
A model for $J/\psi$ absorption in hadronic matter,
\href{https://journals.aps.org/prc/abstract/10.1103/PhysRevC.62.034903}{Phys. Rev. C \textbf{62}, 034903 (2000)}.

\bibitem{Chen:2015igx}
D.~Y.~Chen and Y.~B.~Dong,
Radiative decays of the neutral $Z_c(3900)$,
\href{https://journals.aps.org/prd/abstract/10.1103/PhysRevD.93.014003}{Phys. Rev. D \textbf{93}, 014003 (2016)}.

\bibitem{Becirevic:2009xp}
D.~Becirevic and B.~Haas,
$D^* \to D \pi $ and $D^* \to D \gamma$ decays: Axial coupling and Magnetic moment of $D^*$ meson,
\href{https://link.springer.com/article/10.1140/epjc/s10052-011-1734-y}{Eur. Phys. J. C \textbf{71}, 1734 (2011)}.

\bibitem{BaBar:2008flx}
B.~Aubert \textit{et al.} (\textit{BABAR} Collaboration),
Evidence for $X(3872) \to \psi_{2S} \gamma$ in $B^\pm \to X_{3872} K^\pm$ decays, and a study of $B \to c \bar{c} \gamma K$,
\href{https://journals.aps.org/prl/abstract/10.1103/PhysRevLett.102.132001}{Phys. Rev. Lett. \textbf{102}, 132001 (2009)}.

\bibitem{LHCb:2014jvf}
R.~Aaij \textit{et al.} (LHCb Collaboration),
Evidence for the decay $X(3872)\rightarrow\psi(2S)\gamma$,
\href{https://doi.org/10.1016/j.nuclphysb.2014.06.011}{Nucl. Phys. B\textbf{886}, 665 (2014)}.


\bibitem{BESIII:2020nbj}
M.~Ablikim \textit{et al.} (BESIII Collaboration),
Study of open-charm decays and radiative transitions of the $X(3872)$,
\href{https://journals.aps.org/prl/abstract/10.1103/PhysRevLett.124.242001}{Phys. Rev. Lett. \textbf{124}, 242001 (2020)}.


\bibitem{Cincioglu:2019gzd}
E.~Cincioglu and A.~Ozpineci,
Radiative decay of the $X(3872)$ as a mixed molecule-charmonium state in effective field theory,
\href{https://doi.org/10.1016/j.physletb.2019.134856}{Phys. Lett. B \textbf{797}, 134856 (2019)}.

\bibitem{Gamermann:2007fi}
D.~Gamermann and E.~Oset,
Axial resonances in the open and hidden charm sectors,
\href{https://doi.org/10.1140/epja/i2007-10435-1}{Eur. Phys. J. A \textbf{33} , 119 (2007)}.

\bibitem{Xiao:2019aya}
C.~W.~Xiao, J.~Nieves, and E.~Oset,
Heavy quark spin symmetric molecular states from ${\bar D}^{(*)}\Sigma_c^{(*)}$ and other coupled channels in the light of the recent LHCb pentaquarks,
\href{https://doi.org/10.1103/PhysRevD.100.014021}{Phys. Rev. D \textbf{100}, 014021 (2019) }.

\bibitem{Molina:2020hde}
R.~Molina and E.~Oset,
Molecular picture for the $X_0(2866)$ as a $D^* \bar{K}^*$ $J^P=0^+$ state and related $1^+,2^+$ states,
\href{https://doi.org/10.1016/j.physletb.2020.135870}{Phys. Lett. B \textbf{811}, 135870 (2020);}
[erratum: \href{https://doi.org/10.1016/j.physletb.2022.137645}{\textbf{837}, 137645(E) (2023)}].


\bibitem{Chen:2019asm}
R.~Chen, Z.~F.~Sun, X.~Liu, and S.~L.~Zhu,
Strong LHCb evidence supporting the existence of the hidden-charm molecular pentaquarks,
\href{https://doi.org/10.1103/PhysRevD.100.011502}{Phys. Rev. D \textbf{100}, 011502 (2019) }.

\bibitem{Liu:2020nil}
M.~Z.~Liu, J.~J.~Xie and L.~S.~Geng,
$X_0(2866)$ as a $D^*\bar{K}^*$ molecular state,
\href{https://doi.org/10.1103/PhysRevD.102.091502}{Phys. Rev. D \textbf{102}, 091502 (2020)}.



\bibitem{Ortega:2021yis}
P.~G.~Ortega, D.~R.~Entem, and F.~Fern\'andez,
Does the $J^{PC}=1^{+-}$ counterpart of the $X(3872)$ exist?
\href{https://doi.org/10.1016/j.physletb.2022.137083}{Phys. Lett. B \textbf{829}, 137083 (2022)}.

\bibitem{Lu:2014ina}
J.~X.~Lu, Y.~Zhou, H.~X.~Chen, J.~J.~Xie, and L.~S.~Geng,
Dynamically generated $J^P=1/2^-(3/2^-)$ singly charmed and bottom heavy baryons,
\href{https://journals.aps.org/prd/abstract/10.1103/PhysRevD.92.014036}{Phys. Rev. D \textbf{92}, 014036 (2015)}.


\bibitem{Altenbuchinger:2013vwa}
M.~Altenbuchinger, L.~S.~Geng, and W.~Weise,
Scattering lengths of Nambu-Goldstone bosons off $D$ mesons and dynamically generated heavy-light mesons,
\href{https://journals.aps.org/prd/abstract/10.1103/PhysRevD.89.014026}{Phys. Rev. D \textbf{89}, 014026 (2014)}.





\bibitem{Mutuk:2022ckn}
H.~Mutuk,
Molecular interpretation of $X(3960)$ as $D_s^+ D_s^-$ state,
\href{https://link.springer.com/article/10.1140/epjc/s10052-022-11120-3}{Eur. Phys. J. C \textbf{82}, 1142 (2022)}.

\bibitem{Xin:2022bzt}
Q.~Xin, Z.~G.~Wang, and X.~S.~Yang,
Analysis of the $X(3960)$ and related tetraquark molecular states via the QCD sum rules,
\href{https://link.springer.com/article/10.1007/s43673-022-00070-3}{AAPPS Bull. \textbf{32}, 37 (2022)}.

\bibitem{Bayar:2022dqa}
M.~Bayar, A.~Feijoo, and E.~Oset,
$X(3960)$ seen in $D_s^+D_s^-$ as the X(3930) state seen in $D^+D^-$,
\href{https://doi.org/10.1103/PhysRevD.107.034007}{Phys. Rev. D \textbf{107}, 034007 (2023)}.

\bibitem{Prelovsek:2020eiw}
S.~Prelovsek, S.~Collins, D.~Mohler, M.~Padmanath, and S.~Piemonte,
Charmonium-like resonances with J$^{PC}$ = 0$^{++}$, 2$^{++}$ in coupled $ \mathrm{D}\overline{\mathrm{D}} $, $ {\mathrm{D}}_{\mathrm{s}}{\overline{\mathrm{D}}}_{\mathrm{s}} $ scattering on the lattice,
\href{https://doi.org/10.1007/JHEP06(2021)035}{J.High Energy Phys.06 (2021) 035}.

\bibitem{Liu:2009ei}
X.~Liu and S.~L.~Zhu,
$Y(4143)$ is probably a molecular partner of $Y(3930)$,
\href{https://journals.aps.org/prd/abstract/10.1103/PhysRevD.85.019902}{Phys. Rev. D \textbf{80}, 017502 (2009);}
\href{https://journals.aps.org/prd/abstract/10.1103/PhysRevD.80.017502}{\textbf{85}, 019902 (2012)}.

\bibitem{Wang:2018djr}
E.~Wang, J.~J.~Xie, L.~S.~Geng, and E.~Oset,
The $X(4140)$ and $X(4160)$ resonances in the $e^+e^-\to \gamma J/\psi \phi $ reaction,
\href{https://doi.org/10.1088/1674-1137/43/11/113101}{Chin. Phys. C \textbf{43},  113101 (2019)}.

\bibitem{He:2013oma}
J.~He and P.~L.~L\"u,
Understanding $Y(4274)$ and $X(4320)$ in the $J/\psi \phi$ invariant mass spectrum,
\href{https://doi.org/10.1016/j.nuclphysa.2013.10.001}{Nucl. Phys. A\textbf{919}, 1 (2013)}.

\bibitem{Shen:2010ky}
L.~L.~Shen, X.~L.~Chen, Z.~G.~Luo, P.~Z.~Huang, S.~L.~Zhu, P.~F.~Yu, and X.~Liu,
The molecular systems composed of the charmed mesons in the $H\bar{S}+h.c.$ doublet,
\href{https://link.springer.com/article/10.1140/epjc/s10052-010-1441-0}{Eur. Phys. J. C \textbf{70}, 183 (2010)}.

\bibitem{Finazzo:2011he}
S.~I.~Finazzo, M.~Nielsen, and X.~Liu,
QCD sum rule calculation for the charmonium-like structures in the $J/\psi \phi$ and $J/\psi \omega$ invariant mass spectra,
\href{https://doi.org/10.1016/j.physletb.2011.05.042}{Phys. Lett. B \textbf{701}, 101 (2011)}.

\bibitem{Zhang:2009em}
J.~R.~Zhang and M.~Q.~Huang,
$\{Q \bar{s}\}\{\bar{Q}^{(\prime)}s\}$ molecular states in QCD sum rules,
\href{https://iopscience.iop.org/article/10.1088/0253-6102/54/6/22}{Commun. Theor. Phys. \textbf{54}, 1075 (2010)}.

\bibitem{Zhu:1999ur}
J.~J.~Zhu and T.~N.~Ruan,
Wave functions for massive particles of integer spins,
\href{https://iopscience.iop.org/article/10.1088/0253-6102/32/2/293}{Commun. Theor. Phys. \textbf{32}, 293 (1999)}.

\end{thebibliography}
\end{document}